\documentclass[3p]{elsarticle}
\makeatletter
\def\ps@pprintTitle{%
 \let\@oddhead\@empty
 \let\@evenhead\@empty
 \def\@oddfoot{\centerline{\thepage}}%
 \let\@evenfoot\@oddfoot}
\makeatother

\usepackage{hyperref}
\hypersetup{
    breaklinks=true,   % splits links across lines
    colorlinks=true,   % displays links as colored text instead of blocks
    pdfusetitle=true,  % \title and \author values into pdf metadata etc.
}

\usepackage{amssymb}
\usepackage{graphicx}
\usepackage{amsmath}
\usepackage{placeins}
\usepackage{subfigure}
\usepackage{tikz}
\usepackage{booktabs} % for tables (toprule etc)
\usepackage{multirow}
\usepackage{float} % enforce figure placement (just when writing up the paper)

\usetikzlibrary{shapes.geometric, arrows}
\usetikzlibrary{positioning}

\DeclareMathOperator{\EX}{\mathbb{E}}

\let\emptyset\varnothing

\usepackage{caption}
\usepackage{diagbox} % make 1st cell a diagonal split

\begin{document}

\begin{frontmatter}
\title{Shapley Effect Estimation using Polynomial Chaos}
\author[mysecondaryaddress]{Adrian Stein}
\author[mysecondaryaddress]{Tarunraj Singh}
\cortext[mycorrespondingauthor]{Corresponding author}
\ead{tsingh@buffalo.edu}
\address[mysecondaryaddress]{Department of Mechanical and Aerospace Engineering, University at Buffalo (SUNY),\\ Buffalo, NY 14260-4400, USA}

\begin{abstract}
This paper presents an approach for estimating Shapley effects for use as global sensitivity metrics to quantify the relative importance of uncertain model parameters.  Polynomial Chaos expansion, a well established approach for developing surrogate models is proposed to be used to estimate Shapley effects. Polynomial Chaos permits the transformation of a stochastic process to a deterministic model which can then be used to efficiently evaluate statistical moments of the quantity of interest. These moments include conditional variances which are algebraically mapped to Shapley effects. The polynomial chaos based estimates of Shapley effects are validated using Monte Carlo simulations and tested on the benchmark Ishigami function and on the dynamic SEIR epidemic model and the Bergman Type 1 diabetes model. The results illustrate the correct ranking of uncertain variables for the Ishigami function in contrast to the Sobol indices and illustrates the time-varying rank ordering of the model parameters for the dynamic models.

% Global sensitivity analysis is shown to be a powerful tool to evaluate uncertainty in static or dynamic systems, where the variance-based Sobol analysis is the pioneer. To provide an importance order of input variables of a system, sensitivity indices have to be determined. Two main approaches are used which are either a Monte-Carlo simulation or Polynomial Chaos Expansion. As an alternative method to Sobol analysis it is promising to incorporate the Shapley effect game theory idea into sensitivity analysis. It has been successfully shown that Polynomial Chaos coefficients can map to Sobol indices. This paper tries to provide a method to map those coefficients to Shapley effects and applies the idea on static and dynamic system, when the inputs are independent.
\end{abstract}


\end{frontmatter}
%%%%%%%%%%%%%%%%%%%%%%%%%%%%%%%%%%%%%%%%%%%%%%%%%%%%%%%%
%%%%%%%%%%%%%%%%%%%%%%%%%%%%%%%%%%%%%%%%%%%%%%%%%%%%%%%%
\section{Introduction}
\label{sec:intro}
% A bit of history
Uncertainties are ubiquitous and are broadly classified as Aleatory (irreducible) and Epistemic (reducible). Uncertainty Quantification (UQ) subsumes uncertainty characterization which corresponds to a parametric or non-parametric representation of the uncertainty, and uncertainty propagation which refers to the process of synthesizing the uncertainty in the output due to uncertainties in the input of a model. UQ also includes Global Sensitivity analysis which endeavours to rank the uncertain inputs based on their contributions to the uncertainty in the output. 

Quantifying the uncertainty of systems or processes has gained remarkable interest driven by interest in model reduction, decision making under uncertainty, reliability engineering and explainable machine learning~\cite{Razavi.2021}. Local sensitivity analysis is a decades-old approach which evaluates the variation in a system's output to small perturbations about nominal parameters. Meanwhile, Global Sensitivity Analysis (GSA) endeavours to rank the importance of all input variables over their domain of operation. It should be noted that GSA requires prior knowledge about the distribution of the uncertain input variables. GSA began gaining popularity because of the profound growth of computational power during the end of the last century, where the two most common GSA approaches were (1) Morris method~\cite{Morris.1991} and \mbox{(2) Sobol Analysis}~\cite{Sobol.1993}. The latter, Sobol analysis, is a variance-based analysis and requires the input variables to be independent. 
%Just about the same time the famous Ishigami function~\cite{Ishigami.1990} was created, which has an interesting probability density function (pdf) due to its bi-modality. 
Sobol and Gershman~\cite{Sobol.1995} introduced a derivative-based approach which integrated over the uncertain domain served as the global sensitivity metric. This was 
modified by Kucherenko~\cite{Kucherenko.2009} and called the Derivative-based Global Sensitivity Measures (DGSM). At the same time Sobol' and Kucherenko revealed a relationship between the Sobol indices and the DGSM\cite{Sobol.2009}, which is useful for high dimensional systems~\cite{Iooss.2014}.  Nandi et al.~\cite{Nandi.2019} proposed using a sampling based approach for numerical integration called Conjugate Unscented Transform to more efficiently evaluate Kucherenko's~\cite{Kucherenko.2009} DGSM metrics.

% MC vs PCE
The global sensitivity indices are generally evaluated via Monte Carlo (MC) sampling or quasi-Monte Carlo sampling, and most recently, via Polynomial Chaos Expansion (PCE). The MC makes use of a ``pick-and-freeze'' method \cite{Sobol.1993}\cite{Mara.2021} while a PCE surrogate model can be developed using an intrusive or non-intrusive approach. In the intrusive method, one expands the governing equations using separation of variables and applying the Galerkin projection is called the intrusive method. The non-intrusive approach is a sample based approach and being a black-box approach makes coding rather easy~\cite{Kim.2013}~\cite{Xiu.2002}. 
%and requires more term manipulations than the non-intrusive one. The non-intrusive approach is simply a reconstruction of the output's pdf and figuring out the PC-coefficients via a regression method.
However, for systems with many input variables, it is well know that the PCE can be a computational burden~\cite{Sudret.2008}. An even deeper connection between GSA and PCE was made by Sudret~\cite{Sudret.2008}, who showed that the Sobol indices can be algebraically mapped to the polynomial chaos (PC) -coefficients. Motivated by the computational burden of calculating DGSM measures via MC or quasi MC, Sudret and Mai~\cite{Sudret.2015} found a way to compute DGSM measures with PC-coefficients as well. 

% Applications of PCE
GSA has been applied in disparate domains since the recognition that it should be an integral part of mathematical modeling~\cite{Razavi.2021}. 
Harman and Johnston~\cite{Harman.2016} described in detail how intrusive PCEs are applied on dynamical system, where they chose the Susceptible-Infected-Recovered (SIR) disease model, and compared it to MC results, while performing a Sobol analysis.
A more complex disease model is the Susceptible-Exposed-Infected-Recovered (SEIR) disease model, which was used by Jensen et al.~\cite{Jensen.2022}. Based on Danish data for the spread of Covid-19, they performed a Sobol analysis via PCE. To compute Sobol indices faster, Blatman and Sudret~\cite{Blatman.2010} proposed a sparse PCE which was used by Lin et al.~\cite{Lin.2018} in a 3D multiphysics model for lithium-ion batteries.
In the control engineering field Singh et al.~\cite{Singh.2010} used PCE for the design of robust input shapers. Sobol analysis has also been used in for the analysis of energy consumption of battery electric vehicles~\cite{Braband.2022}, Bayesian networks~\cite{BallesterRipoll.2021}, type 1 diabetes modelling~\cite{Nandi.2020}, state of charge estimation~\cite{Fan.2021} or thermal runaway~\cite{Yeardley.2020} of lithium-ion batteries~\cite{Lai.2021}, underwater gliders~\cite{Wu.2021} or in NOx level sensitivity analysis in premixed burners~\cite{Yousefian.2018}. 

% Borgonovo and Sobol can be wrong
Borgonovo~\cite{Borgonovo.2007} introduced a non-moment based global sensitivity measure which compares the distance between two probability density functions (pdf) as the basis of ranking the relative importance of uncertain input variables. Nandi and Singh~\cite{Nandi.2021} also introduced a variety of non-moment based global sensitivity measures based on statistical distances including the Wasserstein, Hellinger and Kolmogorov distances. They proposed using polynomial chaos surrogate models to alleviate the computational burden of estimating statistical distances. Both of these papers and one by Chun et al.~\cite{Chun.2000} confirmed the incorrect ranking of the three uncertain variables of the Ishigami function generated by the Sobol indices. The variance based Sobol indices orders the variables in descending order of importance as ``1,2,3'', while the non-moment based metrics ranks them as ``2,1,3''. This discrepancy can be attributed to the fact that the Ishigami function is characterized by a bi-modal probability density function which the variance based Sobol metrics are unable to rank order correctly.

% Coming back to the aforementioned Ishigami function, Sobol analysis ranks the three input variables in a ``1,2,3'' order for certain parametric choice, while the Borgonovo metric ranks it as ``2,1,3'', an observation which has been made by Chun et al.~\cite{Chun.2000} earlier with a different metric. Therefore, it became evident that the Sobol analysis can rank the importance of the Ishigami function wrongly because variance-based approaches are not sufficient to capture the bi-modality. Therefore, it is recommended to compare Sobol indices to a pdf sample-based approach like the Borgonovo metric. The failure of ranking, when using Sobol analysis, was nicely shown by Liu et al.~\cite{Liu.2006} with an even simpler example. Additionally to Borgonovo's work~\cite{Borgonovo.2007}, Nandi and Singh~\cite{Nandi.2021} developed non-momentum-based global sensitivity metrics, which are based on statistical distances, where polynomial chaos surrogate models are used to avoid computational cost via MC.

% Shapley
Shapley value is a concept proposed by Lloyd Shapley in 1953~\cite{Shapley.1953} for a fair distribution of credit amongst players based on their value to the coalition, in a cooperative game setting. The Shapley effect quantifies the average marginal contribution of each player by considering all possible coalitions that the player could be a part of. One of the advantage of using Shapley effects as proxies for global sensitivity measures is that the uncertain input variables can be correlated~\cite{Algaba.2019}. The downside of using Shapley effects for a large number of uncertain variables, is the computational cost involved in evaluating the marginal cost of a player in a large selection of coalitions~\cite{Mara.2021}~\cite{Plischke.2021}~\cite{Radaideh.2019}~\cite{Song.2016}~\cite{Iooss.2019}. 

%A different way of looking at GSA is from a cooperative game theory perspective. Shapley introduced the Shapley value theorem~\cite{Shapley.1953}, which addresses Shapley effects to players by looking at their marginal contribution to a certain output. The main advantage of Shapley effects is that the inputs can be correlated~\cite{Algaba.2019} whereas on the other hand they are computationally expensive to calculate, was got noted in several publications such as~\cite{Mara.2021}~\cite{Plischke.2021}~\cite{Radaideh.2019}~\cite{Song.2016}~\cite{Iooss.2017}. 
Owen~\cite{Owen.2014} noted a unique characteristic of the Shapley effects, that is Shapley effects lie between the first-order and total Sobol indices. Iooss and Prieur~\cite{Iooss.2019} while investigating dependent input variables found that the ``sandwich'' effect doesn not always hold true.
%Motivated by the computational burden of calculating Shapley effects, Owen assumed independent input variables~\cite{Owen.2014} and found a ``sandwich'' effect between first-order Sobol indices, Shapley effects and total Sobol indices. 
Owen articulated that Sobol indices are easy to calculate and could be used to bound the Shapley effects for independent variables. 

% Then Owen and Prieur~\cite{Owen.2016} dealt with dependent inputs and compared Shapley effect to the analysis of variance (ANOVA) based sensitivity indices and agreed that the Shapley value is computationally expensive to compute but has useful properties. To reduce the computational cost Plischke et al.~\cite{Plischke.2021} used a M\"obius inverse to compute Shapley effects. 
% %Three years after Owen's finding of the ``sandwich'' effect, Iooss and Prieur~\cite{Iooss.2017} published a paper investigating dependent input variables and found that the ``sandwich'' effect doesn't hold every time. 
% One might say that Sobol indices just hold for independent input variables but with a so called Rosenblatt transformation, which transforms dependent variables to independent ones, such sensitivity indices can be derived when dependencies apply. However, this is not further discussed in this paper because we are dealing with independent inputs. Compared to the Shapley effect definition by Owen~\cite{Owen.2014}, Iooss and Prieur~\cite{Iooss.2017}, Il Idrissi et al.~\cite{IlIdrissi.2021} introduced target Shapley effects and Heredia et al.~\cite{Heredia.2022} aggregated Shapley effects. A way to simplify Shapley's calculation Goda~\cite{Goda.2021} proposed a method to calculate all Shapley effects at the same time or Castro et al.~\cite{Castro.2009} showed a polynomial way of calculating them.

    Shapley effects have recently been used in machine learning for feature selection, model interpretability, and model reduction by pruning~\cite{Rozemberczki.2022}. The domain of explainable or interpretable Artificial Intelligence is exploiting the use of Shapley effects to help comprehend the decisions of outputs of machine learning models since one of the shortcomings of machine learning based models is their opaqueness due to their black box architecture. A classic example of the use of Shapley effects is to allocate landing fees for a proposed runway; based on aircraft size which correlates to length of required runway~\cite{Littlechild.1977}. 
Shapley effects have been used in 
%the Machine Learning domain~\cite{Chau.2022}~\cite{Frye.2021} , 
hydraulic models~\cite{Pheulpin.2022}, biomanufacturing process risk~\cite{Xie.2022}, power flow models for islanded microgrids~\cite{Wang.2018}, drug sensitivity~\cite{Chen.2021}, earth fissures~\cite{Li.2022}, and fatigue blade load estimations~\cite{Schroder.2020}. Some researchers compared Sobol indices and Shapley effects such as:~\cite{Carta.2020},~\cite{Broto.2019},~\cite{Owen.2017},~\cite{Iooss.2019}. Most researchers uses MC sampling to calculate the Shapley effects, such as:~\cite{IlIdrissi.2021}~\cite{Goda.2021}~\cite{Heredia.2022}~\cite{Radaideh.2019}~\cite{Song.2016}~\cite{Wu.2021}, to name a few.
    %Moretti and Patrone provide a list of application which benefit from the use of Shapley effects~\cite{moretti2008transversality}.

% Recently and where our research grabs onto
Recentely, new ideas have entered the GSA community. One of these new ideas, by Gayrard et al.~\cite{Gayrard.2020}, was to perform stochastic processing with independent increments as inputs.  Another, by Song et al.~\cite{Song.2016}, established the connection between the first-order, total, and Shapley effects using the concept of semivalues. Benoumechiara et al.~\cite{Benoumechiara.2019} used a Gaussian Process (Kriging model) to calculate Shapley effects faster than with MC, where they presented their findings using violin plots. With a slightly different parametric choice for the Ishigami function than in~\cite{Chun.2000}~\cite{Borgonovo.2007}, Benoumechiara et al.~\cite{Benoumechiara.2019} and Plischke et al.~\cite{Plischke.2021} ranked the order as ``2,1,3'' derived from Shapley effects. Benoumechiara et al.~\cite{Benoumechiara.2019} acknowledged the relationship between Sobol and Shapley for linear Gaussian models and highlighted Sudret's work~\cite{Sudret.2008} on computing Sobol indices from PC-coefficients. Their final remark states: \textit{It would be interesting to have such a decomposition for the Shapley effects.} because even using Kriging to estimate Shapley effects is computationally expensive, especially in higher dimensions. Our paper is positioned likewise, by establishing a relationship between PC-coefficients and Shapley effects. We will present two static models; which are the Ishigami function in the scheme of~\cite{Benoumechiara.2019}~\cite{Plischke.2021} and a user-defined function with four uncertain input variables. To illustrate the proposed approach on dynamical models, the SEIR disease model and the Bergman type 1 diabetes model are considered. We highlight that this paper is only focused on independent input variables with a hypercube representing the uncertain domain.

% %%% Numeric tools
% For the interested reader we refer to Azzini and Rosati~\cite{Azzini.2022} who list a set of problems which can be used for GSA. As a numerical tool for sensitivity analysis the authors refer to UQLab~\cite{Marelli.2022}. A whole description of how to apply GSA in R can be found in~\cite{Mohammadi.2022}.

%%% Describing the sections
This paper is structured as follows: Section~\ref{sec:variance_based_sensitivity_analysis} describes the variance-based sensitivity analysis, which includes a description of Sobol analysis, Shapley value theorem, PCE, mapping between PC-coefficients, and Sobol indices, as well as the Borgonovo metric. Section~\ref{sec:static_equation} presents the static problems including the Ishigami function and a user-defined function. Section~\ref{sec:dynamic_equation} illustrates the GSA on the SEIR disease model and on the Bergman type 1 diabetes model. Our concluding remarks are presented in section~\ref{sec:conclusion}.
%Another method which can be used in GSA is the Cram\'er-von-Mises distance~\cite{Kala.2020}
%Gamboa et al.~\cite{Gamboa.2020} who introduced new estimators based on Chatterjee coefficient~\cite{Chatterjee.2019} for Cram\'er-von-Mises indices, Sobol indices and Shapley effects.
%%%%%%%%%%%%%%%%%%%%%%%%%%%%%%%%%%%%%%%%%%%%%%%%%%%%%%%%
%%%%%%%%%%%%%%%%%%%%%%%%%%%%%%%%%%%%%%%%%%%%%%%%%%%%%%%%
\section{Review of Global Sensitivity Metrics}
\label{sec:variance_based_sensitivity_analysis}

This section reviews well established global sensitivity metrics including Sobol, a variance based metric and the Borgonovo delta metric, a non-moment based metric. It also proposes the use of the Shapley effect as a metric and illustrates how PC-coefficients can algebraically map to Shapley effects.

%%%%%%%%%%%%%%%%%%%%%%%%%%%%%%%%%%%%%%%%%%%%%%%%%%%%%%%%
\subsection{Sobol indices}
Assume a function is given as:
\begin{align}
    y=f(x), \:\:\: x\in\mathcal{R}^n
\end{align}
where $y$ is a scalar output and $x$ represents the input parameters, presumed to be real numbers. The ANOVA~\cite{Kucherenko.2005} decomposition can be written as~\cite{Sudret.2008} :
\begin{align}
    f(x_1,...,x_n) = f_0 + \sum^{n}_{i=1}f_i(x_i) + \sum_{1\leq i < j \leq n}f_{ij}(x_i,x_j) + ... + f_{1,2,...,n}(x_1,...,x_n)
\end{align}
where $f_0$ is a constant, $f_i$ depends purely on $x_i$ and $f_{i,j}$ purely on $x_i$ and $x_j$. Furthermore, the following condition holds:
\begin{align}
    \int^1_0f_{1,2,...,s}(x_{1}, x_{2},...,x_{s})dx_{k} = 0,\:\: 1 \leq k \leq s
\end{align}
which means that the input variables are independent. It can be stated that:
\begin{subequations}
\begin{align}
    f_o &= \int_\mathcal{R}f(x)dx = \EX\left[y\right] \label{eq:f_o}\\
    f_i(x_i) &= \int_\mathcal{R}f(x)dx_{\sim {i}} - f_0 = \EX\left[y|x_i\right] - f_0\label{eq:f_i}\\
    f_{ij}(x_i,x_j) &= \int_\mathcal{R}f(x)dx_{\sim {i},{j}} = \EX\left[y|x_i,x_j\right] - f_i(x_i) - f_j(x_j) - f_0\label{eq:f_ij}
\end{align}
\end{subequations}
where $\sim {i}$ denotes the integration over all variables except $i$ and implies that $x_i$ is a fixed input variable. It is clear that the total variance of $f(x)$ can be calculated as follows:
\begin{align}
    \mathbb{V} = Var\left(f(x)\right)=\int_\mathcal{R}f^2(x)dx - f_0^2
\end{align}
where it is possible to decompose the total variance as:
\begin{align}
    \mathbb{V} = \sum^n_{i=1}\mathbb{V}_i + \sum_{1\leq i < j \leq n}\mathbb{V}_{ij} + ... + \mathbb{V}_{1,2,...,n}
\end{align}
where we define the set $s$ as $s=1,2,...,n$. Similar to the Eqns.~\eqref{eq:f_o}-\eqref{eq:f_ij} we can represent the variances as:
\begin{align}
\mathbb{V}_i &= \mathbb{V}\left(\EX\left[y|x_i\right]\right)\label{eq:conditional_variance_1}\\
\mathbb{V}_{ij} &= \mathbb{V}\left(\EX\left[y|x_i,x_j\right]\right) - \mathbb{V}_i - \mathbb{V}_j\label{eq:conditional_variance_2}.
\end{align}
The Sobol indices can now be calculated as:
\begin{align}
    S_{i_1,...,i_s} = \frac{\mathbb{V}_{i_1,...i_s}}{\mathbb{V}}
\end{align}
where it is known that
\begin{align}
    \sum_i^nS_i + \sum_{1\leq i < j \leq n} S_{ij} + ... + S_{1,2,...n} = 1.
\end{align}
The total Sobol indices can be calculated as:
\begin{align}
    S_{T_i} = \sum_{\kappa_i}S_{1,...,s}
\end{align}
where $\kappa_i = \{(i_1,...,i_s): \exists k, 1\leq k \leq s, i_k=i\}$. 
%%%%%%%%%%%%%%%%%%%%%%%%%%%%%%%%%%%%%%%%%%%%%%%%%%%%%%%%
\subsection{Shapley value concept}
The Shapley effect of each variable is defined as~\cite{Iooss.2019}:
\begin{align}
    Sh_i &= \frac{1}{d!}\sum_{u\subseteq N \setminus \{i\}} |u|!(d-1-|u|)!\left(c(u\cup \{i\}) - c(u)\right) \label{eq:Shapley_version_1}\\
    \leftrightarrow Sh_i &= \frac{1}{d}\sum_{u\subseteq N \setminus \{i\}} \binom{d-1}{|u|}^{-1} \left(c(u\cup \{i\}) - c(u)\right) \label{eq:Shapley_version_2}
\end{align}
where $d$ is the number of variables, $c(...)$ is the worth or value of the coalition. $c(u \cup \{i\})$ is a coalition including variable $i$ and $c(u)$ excluding variable $i$. Shapley effects are shown as importance measures representing the marginal contribution of variable $i$ to a coalition $u$. The Shapley value concept has the following features~\cite{Iooss.2019}:
\begin{itemize}
    \item Relevant when contribution of each player is unequal
    \item Cooperation among players is beneficial rather than working independently
    \item Shapley effects cannot be negative
    \item Shapley effects sum-up to $1$
\end{itemize}
 These features lead to the assertions that:
\begin{itemize}
    \item Coalition $u$ is at least the empty set $\emptyset$
    \item Examining Eqn.~\eqref{eq:Shapley_version_2}, it is clear that coalition $|u|$ cannot have more than $d-1$ players, because it is already a subset and it is excluding player $i$ from the set, so there can be at most $d-1$ coalitions of that type.
\end{itemize}
Iooss and Prieur~\cite{Iooss.2019} proposed a variance based metric for the worth of a coalition:
\begin{align}
    c(u)=\frac{Var\left(E[Y|X_u]\right)}{Var(Y)} \label{eq:Shapley_c_u}.
\end{align}
By assigning uncertain variables as players in a cooperative game setting, Shapley effects could be used as a Global Sensitivity Metric.
%Then the Shapley concept can be compared to Sobol indices, which is shown in section \ref{sec:static_equation} and \ref{sec:dynamic_equation}. 
% Motivated by pedagogy, a simple example to illustrate the evaluation of Shapley effects is provided in the Appendix. 
%%%%%%%%%%%%%%%%%%%%%%%%%%%%%%%%%%%%%%%%%%%%%%%%%%%%%%%%
\subsection{Polynomial Chaos Expansion}
A PCE has the form of~\cite{Kim.2013}~\cite{Singh.2010}:
\begin{align}
    y(\zeta)=\sum^P_{i=0}a_i\Phi_i(\zeta) \label{eq:PCE}
\end{align}
where $y$ is the output variable, $a_i$ are the PC-coefficients and $\Phi_i$ are the basis functions. $P$ is the degree of the PCE. The orthogonality condition of the basis functions is given by~\cite{Kim.2013}:
\begin{align}
    \left<\Phi_i(\zeta),\Phi_j(\zeta)\right> = \int_\Omega\Phi_i(\zeta)\Phi_j(x)w(\zeta)d\zeta = g_i^2\delta_{ij}  \label{eq:PCE_orthogonality}
\end{align}
where $g_i$ are the coefficients which remain from inner product and $\delta_{ij}$ is the Kronecker delta. $w(x)$ is the pdf of $\zeta$. The Wiener-Askey scheme provides the appropriate polynomial basis functions for standard random variables as illustrated in Table~\ref{table:Wiener_Askey_scheme}.
%%%%%%%%%%%%%%%%%%%%%%%%%%%%
\begin{table}[H]
\caption{Wiener-Askey scheme for polynomial chaos expansion. $k_l$ and $k_u$ are the user-defined lower and upper bounds respectively.}
\centering 
\small
\begin{tabular}{c c c}
\toprule 
Distribution of $\zeta$ & Polynomial family & Support \\
\midrule
Gaussian & Hermite & $\left(-\infty,\infty\right)$\\
gamma & Laguerre & $\left[0,\infty\right)$\\
beta & Jacobi & $\left[k_l,k_u\right]$\\
uniform & Legendre & $\left[k_l,k_u\right]$\\
\bottomrule
\label{table:Wiener_Askey_scheme}
\end{tabular}
\end{table}
%%%%%%%%%%%%%%%%%%%%%%%%%%%%
The PC-coefficients can be determined using an intrusive or non-intrusive approach. The intrusive approach requires evaluation of inner products and might not result in closed form expressions for some nonlinear functions. The non-intrusive approach can be considered a black-box approach which requires solving a least-squares problem to determine the coefficients.
In this paper we use the intrusive approach where the coefficients are determined by exploiting Galerkin projections. Consider Eqn.~\eqref{eq:PCE} which can be approximated as:
\begin{align}
    y(\zeta)=f(\zeta)\approx f_{PC}=\sum^P_{i=0}a_i\Phi_i(\zeta) \label{eq:PCE_v2}.
\end{align}
The left-hand side of Eqn.~\eqref{eq:PCE_v2} represents the equations which are known and need to be approximated with a $P$ order polynomial with coefficients $a_i$ as shown on the right-hand side of Eqn.~\eqref{eq:PCE_v2}.
Assuming $\zeta$ is uniformly distributed and selecting for instance $P=2$ results in the PCE:
\begin{align}
    y(\zeta)=f(\zeta)\approx f_{PC} = a_0\Phi_0(\zeta) + a_1\Phi_1(\zeta) + a_2\Phi_2(\zeta) \label{eq:PCE_v3}.
\end{align}
%Eqn.~\eqref{eq:PCE_v3} can be for instance solved with via the Least squares regression method. In this paper we choose to solve all our system via the Galerkin projection, which is simply a multiplication with $\Phi_i(\zeta)$. 
%Usage of Eqn.~\eqref{eq:PCE_orthogonality} on both sides results in:
The Galerkin projection of Eqn.~\eqref{eq:PCE_v3} results in:
\begin{subequations}
    \begin{align}
        \int f(\zeta) \Phi_0(\zeta) w(\zeta)d\zeta = g_0^2a_0 \label{eq:PCE_a_0}\\
        \int f(\zeta) \Phi_1(\zeta) w(\zeta)d\zeta = g_1^2a_1\label{eq:PCE_a_1}\\
        \int f(\zeta) \Phi_2(\zeta) w(\zeta)d\zeta = g_2^2a_2\label{eq:PCE_a_2}.
    \end{align}
\end{subequations}
From Eqns.~\eqref{eq:PCE_a_0}-\eqref{eq:PCE_a_2}, the PC-coefficients $a_0$ to $a_2$ can be solved by dividing by the constant factors $g_0$ to $g_2$. For problems with multiple random variable, the basis functions are determined by the tensor product of the one-dimensional basis functions as illustrated in Table~\ref{table:PCE_mixed_variables} for two random variables both of which are uniformly distributed.
%The presented principle in Eqns.~\eqref{eq:PCE_a_0}-\eqref{eq:PCE_a_2} can be extended easily to multiple uncertainties in multivariable equations. Table~\ref{table:PCE_mixed_variables} shows how the mixed polynomials would look like if $P=2$ and two uncertain variables are used.
%%%%%%%%%%%%%%%%%%%%%%%%%%%%
\begin{table}[H]
\caption{Polynomials for two uncertain variables (both uniform distributed) in a system for a degree $P=2$.}
\centering 
\small
\begin{tabular}{ c||c|c|c }
\diagbox{$\Phi_i(\zeta_2)$}{$\Phi_i(\zeta_1)$} & $1$ & $\zeta_1$ & $(3\zeta_1^2-1)/2$ \\
\hline\hline
$1$ & $1$ & $\zeta_1$ & $(3\zeta_1^2-1)/2$ \\
\hline
$\zeta_2$ & $\zeta_2$ & $\zeta_1\zeta_2$ &  \\
\hline
$(3\zeta_2^2-1)/2$ & $(3\zeta_2^2-1)/2$ &  & 
\label{table:PCE_mixed_variables}
\end{tabular}
\end{table}
%%%%%%%%%%%%%%%%%%%%%%%%%%%%
The polynomial $\Phi_{ij}$ denotes the polynomial basis function, where $i$ and $j$ are the degrees of the $\zeta_1$ and $\zeta_2$ variables respectively. For this reason, we can write the PCE of the system as:
\begin{align}
   y(\zeta_1,\zeta_2) = f(\zeta_1,\zeta_2)\approx f_{PC} = a_{00}\Phi_{00}(\zeta_1,\zeta_2) + a_{10}\Phi_{10}(\zeta_1,\zeta_2) + a_{01}\Phi_{01}(\zeta_1,\zeta_2) ...\nonumber\\...
   + a_{11}\Phi_{11}(\zeta_1,\zeta_2) + a_{20}\Phi_{20}(\zeta_1,\zeta_2) + a_{02}\Phi_{02}(\zeta_1,\zeta_2).
\end{align}
Multiplying all combinations of $\Phi_{ij}$ with a maximum order of $P=2$ and integrating over $\zeta_1$ and $\zeta_2$ results in $6$ equations in 6 unknowns which permit determining the  PC-coefficients $a_{ij}$.

%%%%%%%%%%%%%%%%%%%%%%%%%%%%%%%%%%%%%%%%%%%%%%%%%%%%%%%%
\subsection{Mapping between PC-coefficients and Sobol Indices}
The mean and variance of the random output expressed via the PCE can be calculated as follows~\cite{Sudret.2008}:
\begin{align}
    \Bar{y_{PC}} &= \EX\left[f_{PC}(\zeta)\right] = \int_\mathcal{R} \sum^P_{i=0}a_i\Phi_i(\zeta)\underbrace{\Phi_0(\zeta)}_{=1}w(\zeta)d\zeta = a_0\\
    \mathbb{V}_{PC} &= Var\left(\sum^P_{i=0}a_i\Phi_i(\zeta)\right) = \int_\mathcal{R}\left(\sum^P_{i=0}a_i\Phi_i(\zeta)\sum^P_{i=0}a_i\Phi_i(\zeta) - a_0^2\right)w(\zeta)d\zeta = \sum^P_{i=1}a_i^2\EX\left[\Phi_i^2(\zeta)\right]
\end{align}
%where the integration bounds and $w(\zeta)$ will always result in $1$. 
Consider a univariate polynomial family and denote it as $C_P(x), P\in \mathbb{N}$, where $P$ is the polynomial degree. $C_P$ could for instance represent the family of Legendre Polynomials. When more than one variable in a system is uncertain, this idea can be generalized to a multivariate polynomial of order $M$ (number of variables) and degree $P$, where the multiplication of univariate polynomials creates the multivariate polynomial of degree $P_m$. However, the resulting degree $P_m$ must be less or equal than $P$. Therefore, we define:
\begin{align}
    \alpha = \{\alpha_i: i=1,...,M\}, \:\:\:\: \alpha_i \geq 0, \:\:\:\: \sum^M_{i=1}\alpha_i \leq P \label{eq:PC_Sobol_mapping_degree_constraint}
\end{align}
which introduces a \textit{set}, where $\alpha_i$ is a natural number and corresponds to the degree of the i-th univariate polynomial with respect to the variable $x_i$. Then, the multivariate polynomial $\Phi_i$ can be expressed as the multiplication of $M$ univariate polynomials:
\begin{align}
   \Phi_i = \Phi_{\alpha}(x_1,...,x_M) = \prod^M_{i=1}\Phi_{\alpha_i}(x_i) \label{eq:PC_Sobol_mapping_Poly_as_set_of_tupels}.
\end{align}
The number of polynomials $n_C$ satisfying Eqn.~\eqref{eq:PC_Sobol_mapping_degree_constraint} is given by~\cite{Sudret.2008}: 
\begin{align}
    n_C = \frac{\left(M+P\right)!}{M!P!}.
\end{align}
Let us define  $\psi_{i_1,...,i_s}$ which consists of $n_C$ $\alpha$-tuples where $i_1,...,i_s$ corresponds to a set of variables which are chosen. Therefore, certain tuple elements are zero.
\begin{align}
    \psi_{i_1,...,i_s} = \forall \alpha:
        \begin{cases}
            \alpha_{k} > 0\:\:\: \text{,when} \:\:\: k\in \{i_1,...,i_s\}\\
            \alpha_{k} = 0\:\:\: \text{,when} \:\:\: k\not\in \{i_1,...,i_s\}
        \end{cases} \label{eq:psi_cases}
\end{align}
where $k=1,...,M$. Eqn.~\eqref{eq:psi_cases} states that any $k$ is non-zero if it lies in $\{i_1,...,i_s\}$. Now, the ANOVA decomposition of $f_{PC}$ can be derived as:
\begin{align}
    f_{PC} = a_0 
    + \sum_{1\leq i_1 \leq M}\sum_{\alpha \in \psi_{i_1}} a_{\alpha} \Phi_{\alpha}(x_{i_1}) 
    + \sum_{1\leq i_1 < i_2\leq M}\sum_{\alpha \in \psi_{i_1,i_2}} a_{\alpha}\Phi_{\alpha}(x_{i_1},x_{i_2}) + ... \nonumber\\
    ... + \sum_{1\leq i_1 < ... < i_s\leq M}\sum_{\alpha \in \psi_{i_1,...,i_s}} a_{\alpha}\Phi_{\alpha}(x_{i_1},...,x_{i_s})
    + ...
    + \sum_{\alpha \in \psi_{1,2,...,M}}a_{\alpha}\Phi_{\alpha}(x_1,...,x_M).
\end{align}
Note that each term $\sum_{\alpha \in \psi_{i_1,...,i_s}}a_{\alpha}\Phi_{\alpha}(x_{i_1},...,x_{i_s})$ reads like: The term $a_{\alpha}\Phi_{\alpha}(x_{i_1},...,x_{i_s})$ gets only considered if the tuple $\alpha$ purely depends on the variables $x_{i_1},...,x_{i_s}$. Accordingly, the PC based Sobol indices can be calculated as~\cite{Sudret.2008}:
\begin{align}
    S_{i_1,...,i_s} = \sum_{\alpha \in \psi_{i_1,...,i_s}}\frac{a^2_\alpha\EX\left[\Phi^2_\alpha\right]}{\mathbb{V}_{PC}}.
\end{align}
Since the PC-coefficients can now be mapped to the Sobol indices, the question arises if the Shapley effects can also be calculated from the PC-coefficients. One could say that we just need to calculate the worth $c(\{...\})$ of all coalitions. For a system with $n$ unknowns, the Sobol indices can mapped to the worth $c(\{...\})$ of all coalitions as follows:
\begin{subequations}
    \begin{align}
        % c(\{i\}) &=  S_i\\
        % c(\{i,j\}) &= S_i + S_j + S_{ij}\\
        % c(\{i,j,k\}) &= S_i + S_j + S_k + S_{ij} + S_{ik} + S_{jk} + S_{ijk}\\
        % & \:\: \vdots\nonumber\\
        % c(\{1,2,..,n\}) &= \sum_{i=0}^n S_i + \sum_{i=0}^n\sum_{j=0}^n S_{ij} + \sum_{i=0}^n\sum_{j=0}^n\sum_{k=0}^n S_{ijk} + \cdots + S_{1,2,..,n}
        % \nonumber\\%
        % \nonumber\\%
        % \nonumber\\%
        c(\{i_1\}) &=  \sum_{\alpha \in \psi_{i_1}}\frac{a^2_\alpha\EX\left[\Phi^2_\alpha\right]}{\mathbb{V}_{PC}}\label{eq:PC_worth_1}\\
        c(\{i_1,i_2\}) &= \sum_{\alpha \subset \psi_{i_1,i_2}}\frac{a^2_\alpha\EX\left[\Phi^2_\alpha\right]}{\mathbb{V}_{PC}} + \sum_{\alpha \in \psi_{i_1,i_2}}\frac{a^2_\alpha\EX\left[\Phi^2_\alpha\right]}{\mathbb{V}_{PC}}\\
        & \:\: \vdots\nonumber\\
        c(\{i_1,...,i_s\}) &= \sum_{\alpha \subset \psi_{i_1,...,i_s}}\frac{a^2_\alpha\EX\left[\Phi^2_\alpha\right]}{\mathbb{V}_{PC}} + \sum_{\alpha \in \psi_{i_1,...,i_s}}\frac{a^2_\alpha\EX\left[\Phi^2_\alpha\right]}{\mathbb{V}_{PC}}\\
        & \:\: \vdots\nonumber\\
        c(\{1,2,..,M\}) &= \sum_{\alpha \subset \psi_{1,2,...,M}}\frac{a^2_\alpha\EX\left[\Phi^2_\alpha\right]}{\mathbb{V}_{PC}} + \sum_{\alpha \in \psi_{1,2,...,M}}\frac{a^2_\alpha\EX\left[\Phi^2_\alpha\right]}{\mathbb{V}_{PC}} \approx 1.\label{eq:PC_worth_4}
    \end{align}
\end{subequations}
Note that $\alpha \subset \psi_{i_1,...,i_s}$ stands for all possible combinations of proper subsets of \mbox{$\{i_1,...,i_s\}$} whereas \mbox{$\alpha \in \psi_{i_1,...,i_s}$} stands for the intersecting set of $\{i_1,...,i_s\}$. For a set $\{i_1,...,i_s\}$ we can derive $2^s-1$ proper subsets, because the empty set is a proper subset of every set except for the empty set itself. The worths presented in Eqns.~\eqref{eq:PC_worth_1}-\eqref{eq:PC_worth_4} can be directly substituted into Eqn.~\eqref{eq:Shapley_version_1}.

\begin{figure}[H]
\centering
	\includegraphics[width=0.5\textwidth]{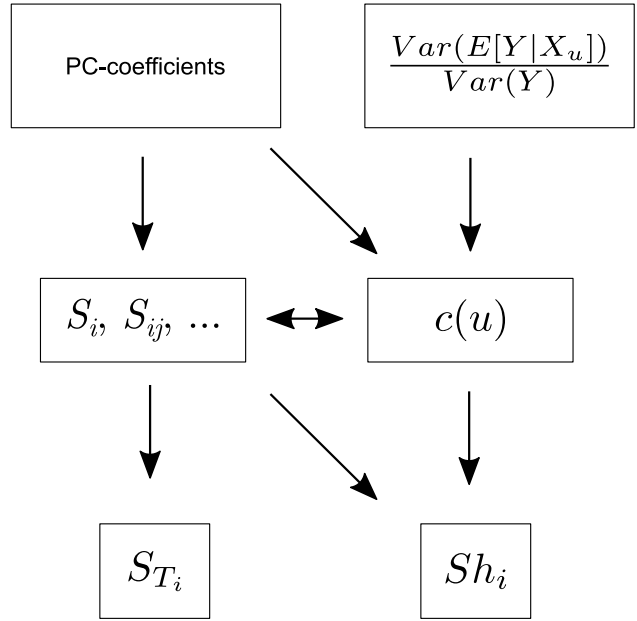}
	\caption{Connection between Sobol indices and Shapley effects.}
 \label{fig:Sobol_Shapley_linkage}
\end{figure}

If one  maps PC-coefficients to Sobol indices, the Shapley effects can be calculated as:
\begin{align}
    Sh_i = S_i + \frac{1}{2} \sum_{1 \leq i < j \leq M}S_{ij} + \frac{1}{3} \sum_{1 \leq i < j < k \leq M}S_{ijk} + ... +  \frac{1}{M-1} \sum_{1 \leq i < ... < s \leq M}S_{i,...,s} +  \frac{1}{M} S_{1,2,...,M}.
\end{align}
Figure~\ref{fig:Sobol_Shapley_linkage} shows the difference between calculating Shapley effects with Eqn.~\eqref{eq:Shapley_c_u} and PC-coefficients. This sheds light on the relation of Sobol indices and Shapley effects. Moreover given all values of $c(u)$, it is possible to map to the corresponding $S_i, S_{ij}, \hdots$ and vice versa.
%%%%%%%%%%%%%%%%%%%%%%%%%%%%%%%
%%%%%%%%%%%%%%%%%%%%%%%%%%%%%%%
%%%%%%%%%%%%%%%%%%%%%%%%%%%%%%%%%%%%%%%%%%%%%%%%%%%%%%%%
\subsection{Borgonovo metric}
\label{subsec:Borgonovo_index}
\begin{figure}[H]
\centering
	\includegraphics[width=0.45\textwidth]{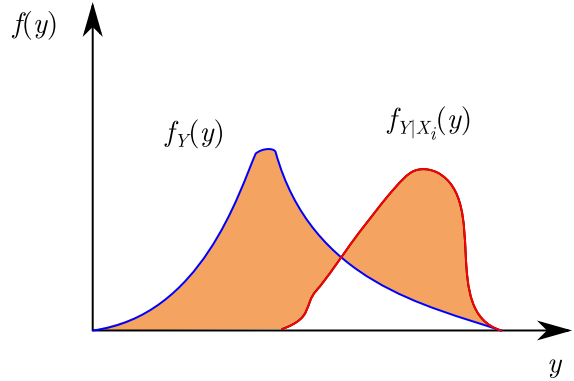}
	\caption{The shift is the area of difference between the two probability density functions (shaded region).}
 \label{fig:Borgonovo_index}
\end{figure}

The Borgonovo metric~\cite{Borgonovo.2007} is a method which measures the shift between the unconditional and conditional pdfs, which is illustrated in Figure~\ref{fig:Borgonovo_index}.
%%%%%%%%%%%%%%%%%%%%%%%%%%%%%%%
%%%%%%%%%%%%%%%%%%%%%%%%%%%%%%%
Assume that a function $Y$ has $i$ uncertain variables. The shift can then be used to provide a ranking order of importance for all uncertain variables. The main idea is that variables with little impact on the output $Y$ show a small shift and variables with a large impact have a large shift. The definition of a shift, which is noted as $Z$, between two pdfs can be written as~\cite{Borgonovo.2007}:
\begin{align}
    Z(X_i) = \int \lvert f_Y(y)-f_{Y|X_i}(y) \rvert dy
\end{align}
where $f_Y(y)$ is the pdf of the output $Y$ and $f_{Y|X_i}(y)$ is the conditional pdf for a given variable $x_i$.  Furthermore, the expected shift between $f_Y(y)$ and $f_{Y|X_i}(y)$ can be written as:
\begin{align}
    E_{X_i}\left[Z(X_i)\right] = \int f_{X_i}(x_i) \left( \int \lvert f_Y(y)-f_{Y|X_i}(y) \rvert dy \right) dx_i.
\end{align}
Finally, the Borgonovo index~\cite{Borgonovo.2007} can be derived from the expected shift, which is:
\begin{align}
    \delta_i = \frac{1}{2} E_{X_i}\left[Z(X_i)\right]
\end{align}
where $\delta_i$ is a moment independent sensitivity measurement for an uncertain variable $x_i$ with respect to an output $Y$.
%%%%%%%%%%%%%%%%%%%%%%%%%%%%%%%%%%%%%%%%%%%%%%%%%%%%%%%%
%%%%%%%%%%%%%%%%%%%%%%%%%%%%%%%%%%%%%%%%%%%%%%%%%%%%%%%%
\section{Algebraic Examples}
\label{sec:static_equation}
This section will present two algebraic examples including the Ishigami function and a user-defined function.
%%%%%%%%%%%%%%%%%%%%%%%%%%%%%%%%%%%%%%%%%%%%%%%%%%%%%%%%
\subsection{Ishigami function}
\label{subsec:Ishigami}
\begin{figure}[H]
\centering
	\includegraphics[width=0.55\textwidth]{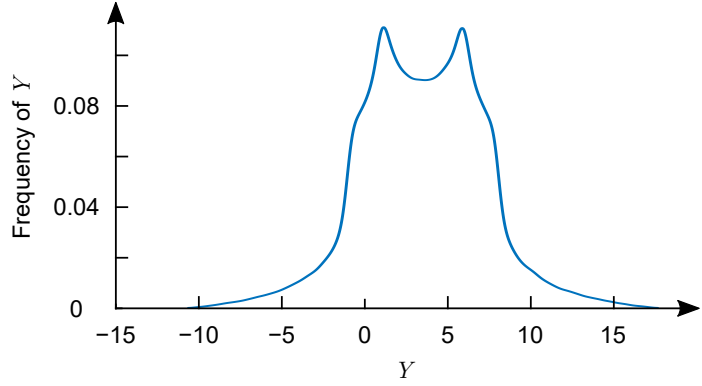}
	\caption{Probability density function of the Ishigami function when $a=7$ and $b=0.1$.}
 \label{fig:Ishigami}
\end{figure}

A benchmark problem for global sensitivity analysis is the Ishigami function, especially because of its bi-modal pdf. The function is given as:
\begin{align}
    Y &=\sin(x_1) + a\sin(x_2)^2 + bx_3^4\sin(x_1) \label{eq:Ishigami_Y}
\end{align}
where $a=7$ and $b=0.1$ are parameters consistent with those used in the literature. $x_1$, $x_2$ and $x_3$ are uncertain variables, where it is assumed that all variables are uniformly distributed between $-\pi$ and $\pi$. The pdf of the Ishigami function is shown in Figure~\ref{fig:Ishigami}.
%%%%%%%%%%%%%%%%%%%%%%%%%%%%%%%
%%%%%%%%%%%%%%%%%%%%%%%%%%%%%%%
The expected value and the total variance can be analytically evaluated by integrating over all three uncertain variables:
\begin{align}
    E[Y] &=\mu= \left(\frac{1}{2\pi}\right)^3 \int^\pi_{-\pi}\int^\pi_{-\pi}\int^\pi_{-\pi}\sin(x_1) + 7\sin(x_2)^2 + 0.1x_3^4\sin(x_1) dx_1dx_2dx_3\\
    Var(Y) &= \left(\frac{1}{2\pi}\right)^3 \int^\pi_{-\pi}\int^\pi_{-\pi}\int^\pi_{-\pi} \left(\sin(x_1) + 7\sin(x_2)^2 + 0.1x_3^4\sin(x_1) -\mu\right)^2dx_1dx_2dx_3.
\end{align}
To determine the Sobol indices, three first-order, three second-order and one third-order indices need to be computed. 
%To calculate the Sobol indices the variances of conditional expectations are needed. 
For the first-order Sobol index, the conditional variances can be evaluated as:
\begin{align}
    E[Y|x_i] &= \mu_i =\left(\frac{1}{2\pi}\right)^2\int^\pi_{-\pi}\int^\pi_{-\pi}\sin(x_1) + 7\sin(x_2)^2 + 0.1x_3^4\sin(x_1)d\mathcal{X}_{\sim i}\\
    Var(E[Y|x_i]) &= \frac{1}{2\pi}\int^\pi_{-\pi}\left(\mu_i - \frac{1}{2\pi}\int^\pi_{-\pi}\mu_idx_i\right)^2dx_i\\
    \rightarrow \mathbb{V}_{i} &=  Var(E[Y|x_{i}])
\end{align}
where $d\mathcal{X}_{\sim i}$ means that the integral is calculated for all variables except $x_i$. For the second-order Sobol indices, the conditional variances can be specifically calculated as:
\begin{align}
    E[Y|x_{ij}] &= \mu_{ij} = \frac{1}{2\pi}\int^\pi_{-\pi}\sin(x_1) + 7\sin(x_2)^2 + 0.1x_3^4\sin(x_1)d\mathcal{X}_{\sim i,j}\\
    Var(E[Y|x_{ij}]) &= \left(\frac{1}{2\pi}\right)^2\int^\pi_{-\pi}\int^\pi_{-\pi}\left(\mu_{ij} - \left(\frac{1}{2\pi}\right)^2\int^\pi_{-\pi}\int^\pi_{-\pi}\mu_{ij}dx_idx_j\right)^2dx_idx_j\\
    \rightarrow \mathbb{V}_{ij} &=  Var(E[Y|x_{ij}]) - \mathbb{V}_i - \mathbb{V}_j.
\end{align}
For the third-order Sobol the conditional variances can be specifically calculated as:
\begin{align}
    E[Y|x_{123}] &= \mu_{123} = \sin(x_1) + 7\sin(x_2)^2 + 0.1x_3^4\sin(x_1)\\
    Var(E[Y|x_{123}]) &= \left(\frac{1}{2\pi}\right)^2\int^\pi_{-\pi}\int^\pi_{-\pi}\int^\pi_{-\pi}\left(\mu_{123} - \left(\frac{1}{2\pi}\right)^2\int^\pi_{-\pi}\int^\pi_{-\pi}\int^\pi_{-\pi}\mu_{123}dx_1dx_2dx_3\right)^2dx_1dx_2dx_3\\
    \rightarrow \mathbb{V}_{123} &=  Var(E[Y|x_{123}]) - \mathbb{V}_{12} - \mathbb{V}_{13} - \mathbb{V}_{23} - \mathbb{V}_1 - \mathbb{V}_2 - \mathbb{V}_3.
\end{align}
The Sobol indices from the analytical evaluation are presented in Table~\ref{table:Ishigami_Sobol_indices_analytical_and_PC} to permit their comparison to those determined using the PCE surrogate model.
%%%%%%%%%%%%%%%%%%%%%%%%%%%%
\begin{table}[H]
\caption{Sobol indices of the Ishigami function.}
\centering 
\small
\begin{tabular}{l l l l l l l}
\toprule 
 Index & Analytical & \multicolumn{4}{c}{PCE order}\\
 & & $P=3$ & $P=5$ & $P=7$ & $P=9$\\
 \midrule
 \midrule
 $S_1$ & 0.3139 & 0.6582 & 0.3699 & 0.3166 & 0.3140\\
 $S_2$ & 0.4424 & 0.0540 & 0.3545 & 0.4377 & 0.4423\\
 $S_3$ & 0 & 0 & 0 & 0 & 0\\
 $S_{12}$ & 0 & 0 & 0 & 0 & 0\\
 $S_{13}$ & 0.2437 & 0.2878 & 0.2756 & 0.2456 & 0.2437\\
 $S_{23}$ & 0 & 0 & 0 & 0 & 0\\
 $S_{123}$ & 0 & 0 & 0 & 0 & 0\\
 \midrule
 $S_{T,1}$ & 0.5576 & 0.9460 & 0.6455 & 0.5623 & 0.5577\\
 $S_{T,2}$ & 0.4424 & 0.0540 & 0.3545 & 0.4377 & 0.4423\\
 $S_{T,3}$ & 0.2437 & 0.2878 & 0.2756 & 0.2456 & 0.2437\\
 \bottomrule
 \label{table:Ishigami_Sobol_indices_analytical_and_PC}
 \end{tabular}
 \end{table}
%%%%%%%%%%%%%%%%%%%%%%%%%%%%
Table~\ref{table:Ishigami_Sobol_indices_analytical_and_PC} shows that, with increasing PC degree, the Sobol indices converge to the analytical solution. 
%However, it is possible to compute Sobol indices and Shapley effects in closed form. 
To determine the Shapley effects, the worth of a coalition can also be calculated analytically with Eqn.~\eqref{eq:Shapley_c_u} and is presented in Table~\ref{table:Ishigami_Shapley_values_analytical_and_PC}.
%%%%%%%%%%%%%%%%%%%%%%%%%%%%
\begin{table}[H]
\caption{Shapley effects of the Ishigami function.}
\centering 
\small
\begin{tabular}{l l l l l l l}
\toprule 
Worth & Analytical & \multicolumn{4}{c}{PCE order}\\
 & & $P=3$ & $P=5$ & $P=7$ & $P=9$\\
 \midrule
 \midrule
 $c(\{1\})$ & 0.3139 & 0.6582 & 0.3699 & 0.3166 & 0.3140\\
 $c(\{2\})$ & 0.4424 & 0.0540 & 0.3545 & 0.4377 & 0.4423\\
 $c(\{3\})$ & 0 & 0 & 0 & 0 & 0\\
 $c(\{1,2\})$ & 0.7563 & 0.7122 & 0.7244 & 0.7544 & 0.7563\\
 $c(\{1,3\})$ & 0.5576 & 0.9460 & 0.6455 & 0.5623 & 0.5577\\
 $c(\{2,3\})$ & 0.4424 & 0.0540 & 0.3545 & 0.4377 & 0.4423\\
 $c(\{1,2,3\})$ & 1 & 1 & 1 & 1 & 1\\
 \midrule
 $Sh_1$ & 0.4357 & 0.8021 & 0.5077 & 0.4395 & 0.4358\\
 $Sh_2$ & 0.4424 & 0.0540 & 0.3545 & 0.4377 & 0.4423\\
 $Sh_3$ & 0.1218 & 0.1439 & 0.1378 & 0.1228 & 0.1219\\
 \bottomrule
 \label{table:Ishigami_Shapley_values_analytical_and_PC}
 \end{tabular}
 \end{table}
%%%%%%%%%%%%%%%%%%%%%%%%%%%%
Note that the worth of an empty coalition is $c(\{\emptyset\})=0$. Similar to the Sobol indices in Table~\ref{table:Ishigami_Sobol_indices_analytical_and_PC}, Table~\ref{table:Ishigami_Shapley_values_analytical_and_PC} illustrates the convergence of the coalitions worth to their analytical values when the PC degree $P$ is increased and therefore a convergence of the Shapley effects to the analytical solution can also be noted. The remaining questions is: What exactly makes computing Shapley effects different from the Sobol analysis? Sobol and Shapley can be used to rank the importance of uncertain variables in a system. This means that a higher ranked uncertain variable contributes more to the perturbation of the output variable. Taking the route of computing Shapley effects via the Sobol indices from Figure~\ref{fig:Sobol_Shapley_linkage}, we can try to look at the weighting of those indices. Table~\ref{table:Ishigami_Sobol_vs_Shapley_analytical_weights} is specifically tailored for the Ishigami function and provides a deeper understanding of the weighting.
%%%%%%%%%%%%%%%%%%%%%%%%%%%%
\begin{table}[H]
\caption{Difference in weighting the Sobol indices when calculating total Sobol indices and Shapley effects for the Ishigami function.}
\centering 
\small
\begin{tabular}{l l l l}
\toprule 
 Sobol Index & Weight for $S_{T}$ & Weight for $Sh$ & Analytical\\
 \midrule
 \midrule
 $S_1$ & 1 & 1 & 0.3139\\
 $S_2$ & 1 & 1 & 0.4424\\
 $S_3$ & 1 & 1 & 0\\
 $S_{12}$ & 1 & 1/2 & 0\\
 $S_{13}$ & 1 & 1/2 & 0.2437\\
 $S_{23}$ & 1 & 1/2 & 0\\
 $S_{123}$ & 1 & 1/3 & 0\\
 \bottomrule
 \label{table:Ishigami_Sobol_vs_Shapley_analytical_weights}
 \end{tabular}
 \end{table}
%%%%%%%%%%%%%%%%%%%%%%%%%%%%
 If we use the analytical values from Table~\ref{table:Ishigami_Sobol_vs_Shapley_analytical_weights} we can calculate the Shapley effects from the Sobol indices as follows:
\begin{subequations}
    \begin{align}
        Sh_1 = S_1 + \frac{1}{2}S_{12} + \frac{1}{2}S_{13} + \frac{1}{3}S_{123} = 0.4357 \label{eq:Ishigami_Sh_1_analytical}\\
        Sh_2 = S_2 + \frac{1}{2}S_{12} + \frac{1}{2}S_{23} + \frac{1}{3}S_{123}
        = 0.4424 \label{eq:Ishigami_Sh_2_analytical}\\
        Sh_3 = S_3 + \frac{1}{2}S_{13} + \frac{1}{2}S_{23} + \frac{1}{3}S_{123}
        = 0.1218 \label{eq:Ishigami_Sh_3_analytical}.
    \end{align}
\end{subequations}
%A detailed calculation how Eqn.~\eqref{eq:Shapley_version_2} is applied to the 3 variable Ishigami problem is illustrated in the Appendix. 
Conversely, the total Sobol index can be calculated as follows:
\begin{subequations}
    \begin{align}
        S_{T,1} = S_1 + S_{12} + S_{13} + S_{123} = 0.5576 \label{eq:Ishigami_ST_1_analytical}\\
        S_{T,2} = S_2 + S_{12} + S_{23} + S_{123} = 0.4424 \label{eq:Ishigami_ST_2_analytical}\\
        S_{T,3} = S_3 + S_{13} + S_{23} + S_{123} = 0.2437 \label{eq:Ishigami_ST_3_analytical}.
    \end{align}
\end{subequations}
The results from Eqns.~\eqref{eq:Ishigami_Sh_1_analytical}-\eqref{eq:Ishigami_Sh_3_analytical} compared to Eqns.~\eqref{eq:Ishigami_ST_1_analytical}-\eqref{eq:Ishigami_ST_3_analytical} show that Shapley is weighing the higher order Sobol indices lower compared to Sobol where each index is equally weighted. Evaluating the ranking from a Sobol perspective suggests to order the uncertain variables as $x_1$, $x_2$ and $x_3$, while Shapley suggests $x_2$, $x_1$ and $x_3$. In fact, we can note if only first-order Sobol indices exist and higher order ones are zero, the total Sobol indices will be identical with the Shapley effects.

As pointed out in \cite{Plischke.2021} and \cite{Benoumechiara.2019}, the uncertain variable $x_2$ is more important than $x_1$. The clearest way of evaluating the ranking of the uncertain variables is to compute the Borgonovo index as described in subsection~\ref{subsec:Borgonovo_index} or other statistical distances~\cite{Nandi.2021} because they are based on pdfs which can be computational expensive. $300,000$ MC evaluations to calculate the Borgonovo index reveals the uncertain variables are ranked as $x_2$, $x_1$ and $x_3$, which is the same rank order which Shapley suggests. Results of all three methods are listed in Table~\ref{table:Ishigami_Sobol_indices_and_Shapley_values_comparision}.
%%%%%%%%%%%%%%%%%%%%%%%%%%%%
\begin{table}[H]
\caption{Total Sobol indices, Shapley effects and Borgonovo index for the Ishigami function.}
\centering 
\small
\begin{tabular}{l c c c}
\toprule 
 Variable & Total Sobol & Shapley & Borgonovo\\
 \midrule
 \midrule
 $x_1$ & 0.5576 & 0.4357 & 0.2392\\
 $x_2$ & 0.4424 & 0.4424 & 0.4222\\
 $x_3$ & 0.2437 & 0.1218 & 0.1957\\
 \bottomrule
 \label{table:Ishigami_Sobol_indices_and_Shapley_values_comparision}
 \end{tabular}
 \end{table}
%%%%%%%%%%%%%%%%%%%%%%%%%%%%
%%%%%%%%%%%%%%%%%%%%%%%%%%%%%%%%%%%%%%%%%%%%%%%%%%%%%%%%
\subsection{User-defined function}
\label{subsec:user_defined}
\begin{figure}[H]
\centering
	\includegraphics[width=0.55\textwidth]{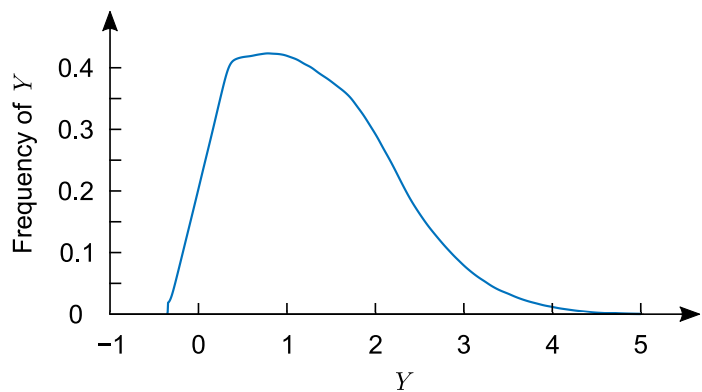}
	\caption{Probability density function of the user-defined function.}
 \label{fig:User_defined}
\end{figure}

We consider a polynomial function to illustrate the evaluation of the Shapley effects where:
\begin{align}
    Y = 1.9x_1^2 + 2x_2^2 + 1.05x_3x_1^3 + 0.35x_4
\end{align}
where it is assumed that $x_1$ to $x_4$ are uniformly distributed and each variable lies between $-1$ and $1$. Figure~\ref{fig:User_defined} illustrates the pdf of the user-defined function. 
%%%%%%%%%%%%%%%%%%%%%%%%%%%%%%%
%%%%%%%%%%%%%%%%%%%%%%%%%%%%%%%
The Sobol indices are listed in Table~\ref{table:User_defined_Sobol_indices_analytical_and_PC} where it is evident that for a PCE order of $P=4$, we obtain the exact analytical solution. Since the highest order of the user-defined function is $4$, a PCE of order $4$ can exactly reproduce the original function. Note that all higher order Sobol indices are $0$ except $S_{13}$.
%%%%%%%%%%%%%%%%%%%%%%%%%%%%
\begin{table}[H]
\caption{Sobol indices of the user-defined function.}
\centering 
\small
\begin{tabular}{l l l l l l l}
\toprule 
 Index & Analytical & \multicolumn{4}{c}{PCE order}\\
 & & $P=1$ & $P=2$ & $P=3$ & $P=4$\\
 \midrule
 \midrule
 $S_1$ & 0.4169 & 0 & 0.4215 & 0.4215 & 0.4169\\
 $S_2$ & 0.4619 & 0 & 0.4670 & 0.4670 & 0.4169\\
 $S_3$ & 0 & 0 & 0 & 0 & 0\\
 $S_4$ & 0.0530 & 1 & 0.0536 & 0.0536 & 0.0530\\
 $S_{12}$ & 0 & 0 & 0 & 0 & 0\\
 $S_{13}$ & 0.0682 & 0 & 0.0579 & 0 & 0.0682\\
 $S_{14}$ & 0 & 0 & 0 & 0 & 0\\
 $S_{23}$ & 0 & 0 & 0 & 0 & 0\\
 $S_{24}$ & 0 & 0 & 0 & 0 & 0\\
 $S_{34}$ & 0 & 0 & 0 & 0 & 0\\
 $S_{123}$ & 0 & 0 & 0 & 0 & 0\\
 $S_{124}$ & 0 & 0 & 0 & 0 & 0\\
 $S_{134}$ & 0 & 0 & 0 & 0 & 0\\
 $S_{234}$ & 0 & 0 & 0 & 0 & 0\\
 $S_{1234}$ & 0 & 0 & 0 & 0 & 0\\
 \midrule
 $S_{T,1}$ & 0.4851 & 0 & 0.4794 & 0.4794 & 0.4851\\
 $S_{T,2}$ & 0.4619 & 0 & 0.4670 & 0.4670 & 0.4619\\
 $S_{T,3}$ & 0.0682 & 0 & 0.0579 & 0.0579 & 0.0682\\
 $S_{T,4}$ & 0.0530 & 1 & 0.0536 & 0.0536 & 0.0530\\
 \bottomrule
 \label{table:User_defined_Sobol_indices_analytical_and_PC}
 \end{tabular}
 \end{table}
%%%%%%%%%%%%%%%%%%%%%%%%%%%%
Table~\ref{table:User_defined_Shapley_values_analytical_and_PC} illustrates the convergence of the coalition worths and Shapley effects when higher order PCE  are used. Again, for a PCE order of $P=4$, we can observe that the PCE approximation is identical to the analytical results.
%%%%%%%%%%%%%%%%%%%%%%%%%%%%
\begin{table}[H]
\caption{Shapley effects of the user-defined function.}
\centering 
\small
\begin{tabular}{l l l l l l l}
\toprule 
 Worth & Analytical & \multicolumn{4}{c}{PCE order}\\
 & & $P=1$ & $P=2$ & $P=3$ & $P=4$\\
 \midrule
 \midrule
 $c(\{1\})$ & 0.4169 & 0 & 0.4215 & 0.4215 & 0.4169\\
 $c(\{2\})$ & 0.4619 & 0 & 0.4670 & 0.4670 & 0.4619\\
 $c(\{3\})$ & 0 & 0 & 0 & 0 & 0\\
 $c(\{4\})$ & 0.0530 & 1 & 0.0536 & 0.0536 & 0.0530\\
 $c(\{1,2\})$ & 0.8788 & 0 & 0.8884 & 0.8884 & 0.8788\\
 $c(\{1,3\})$ & 0.4851 & 0 & 0.4794 & 0.4794 & 0.4851\\
 $c(\{1,4\})$ & 0.4699 & 1 & 0.4751 & 0.4751 & 0.4699\\
 $c(\{2,3\})$ & 0.4619 & 0 & 0.4670 & 0.4670 & 0.4619\\
 $c(\{2,4\})$ & 0.5149 & 1 & 0.5206 & 0.5206 & 0.5149\\
 $c(\{3,4\})$ & 0.0530 & 1 & 0.0536 & 0.0536 & 0.0530\\
 $c(\{1,2,3\})$ & 0.9470 & 0 & 0.9464 & 0.9464 & 0.9470\\
 $c(\{1,2,4\})$ & 0.9318 & 1 & 0.9421 & 0.9421 & 0.9318\\
 $c(\{1,3,4\})$ & 0.5381 & 1 & 0.5330 & 0.5330 & 0.5381\\
 $c(\{2,3,4\})$ & 0.5149 & 1 & 0.5206 & 0.5206 & 0.5149\\
 $c(\{1,2,3,4\})$ & 1 & 1 & 1 & 1 & 1\\
 \midrule
 $Sh_1$ & 0.4510 & 0 & 0.4504 & 0.4504 & 0.4510\\
 $Sh_2$ & 0.4619 & 0 & 0.4670 & 0.4670 & 0.4619\\
 $Sh_3$ & 0.0341 & 0 & 0.0290 & 0.0290 & 0.0341\\
 $Sh_4$ & 0.0530 & 1 & 0.0536 & 0.0536 & 0.0530\\
 \bottomrule
 \label{table:User_defined_Shapley_values_analytical_and_PC}
 \end{tabular}
 \end{table}
%%%%%%%%%%%%%%%%%%%%%%%%%%%%
Evaluating the ranking from a Sobol perspective suggests to order the uncertain variables as $x_1$, $x_2$, $x_3$ and $x_4$, while Shapley suggests $x_2$, $x_1$, $x_4$ and $x_3$. Table~\ref{table:User_defined_Sobol_indices_and_Shapley_values_comparision} provides the Borgonovo indices, where the ranking is identical to the ranking based on the Shapley effects.
%%%%%%%%%%%%%%%%%%%%%%%%%%%%
\begin{table}[H]
\caption{Total Sobol indices, Shapley effects and Borgonovo index for the user-defined function.}
\centering 
\small
\begin{tabular}{l c c c}
\toprule 
 Variable & Total Sobol & Shapley  & Borgonovo\\
 \midrule
 \midrule
 $x_1$ & 0.4851 & 0.4510 & 0.2734\\
 $x_2$ & 0.4619 & 0.4619 & 0.3309\\
 $x_3$ & 0.0682 & 0.0341 & 0.0178\\
 $x_4$ & 0.0530 & 0.0530 & 0.0800\\
 \bottomrule
 \label{table:User_defined_Sobol_indices_and_Shapley_values_comparision}
 \end{tabular}
 \end{table}
%%%%%%%%%%%%%%%%%%%%%%%%%%%%
%%%%%%%%%%%%%%%%%%%%%%%%%%%%%%%%%%%%%%%%%%%%%%%%%%%%%%%%
%%%%%%%%%%%%%%%%%%%%%%%%%%%%%%%%%%%%%%%%%%%%%%%%%%%%%%%%
\section{Dynamic Systems}
\label{sec:dynamic_equation}

Global sensitivity analysis of dynamic models is necessary for control and forecasting problems. 
%Especially for control problems dynamical system are more important than static equations. 
Therefore, for illustrative purposes, this section will present two models including a SEIR disease model~\cite{Jensen.2022} and the Bergman model, which is used to model Type 1 diabetes.
%%%%%%%%%%%%%%%%%%%%%%%%%%%%%%%%%%%%%%%%%%%%%%%%%%%%%%%%
\subsection{SEIR model}
The nonlinear SEIR model is presented by the equations:
\begin{subequations}
    \begin{align}
        \frac{\partial S}{\partial t} &= -\beta\frac{I(t)S(t)}{N}\label{eq:SEIR_original_S}\\
        \frac{\partial E}{\partial t} &= \beta\frac{I(t)S(t)}{N} - \sigma E(t)\label{eq:SEIR_original_E}\\
        \frac{\partial I}{\partial t} &= \sigma E(t) - \gamma I(t)\label{eq:SEIR_original_I}\\
        \frac{\partial R}{\partial t} &= \gamma I(t)\label{eq:SEIR_original_R}
    \end{align}
\end{subequations}
where $\beta$, $\sigma$ and $\gamma$ are assumed to be uncertain. $S$, $E$, $I$ and $R$, stand for the susceptible, exposed, infected and recovered group respectively. Variable $N$ is the total population and is set to $1$ for simplicity. The uncertain parameters can be interpreted as follows:
\begin{itemize}
    \item $\gamma = \frac{1}{\tau_{inf}}$, where $\tau_{inf}$ is the time constant of the infectious state
    \item $\sigma = \frac{1}{\tau_{inc}}$, where $\tau_{inc}$ is the time constant of the exposed state
    \item $\beta = R_o \gamma$, where $R_o$ is the reproduction number
\end{itemize}
We assume that they are uniformly distributed as follows \cite{Harman.2016}:
\begin{subequations}
    \begin{align}
        \beta \sim U(2.5, 5.5)\\
        \sigma \sim U(0.5, 1.5)\\
        \gamma \sim U(0.5, 1.5)
    \end{align}
\end{subequations}
which can be further rewritten as:
\begin{subequations}
    \begin{align}
        \beta = \beta_0 + \beta_1 \zeta_1\\
        \sigma = \sigma_0 + \sigma_1 \zeta_2\\
        \gamma = \gamma_0 + \gamma_1 \zeta_3
    \end{align}
\end{subequations}
where the subscript $(.)_0$ is the mean and $(.)_1$ is used to map the uniform distribution with arbitrary bounds to the interval $[ -1, \: 1]$. PCE of the random SEIR variables are:
\begin{subequations}
\begin{align}
    S(t,\zeta_1,\zeta_2,\zeta_3) &= \sum^{\infty}_{i=0}S_i(t)\Phi(\zeta_1,\zeta_2,\zeta_3) = \sum^{\infty}_{i=0}\sum^{\infty}_{j=0}\sum^{\infty}_{k=0}S_{ijk}(t)\Phi_i(\zeta_1)\Phi_j(\zeta_2)\Phi_k(\zeta_3)\label{eq:SEIR_S}\\
    E(t,\zeta_1,\zeta_2,\zeta_3) &= \sum^{\infty}_{i=0}E_i(t)\Phi(\zeta_1,\zeta_2,\zeta_3) =
    \sum^{\infty}_{i=0}\sum^{\infty}_{j=0}\sum^{\infty}_{k=0}E_{ijk}(t)\Phi_i(\zeta_1)\Phi_j(\zeta_2)\Phi_k(\zeta_3)\label{eq:SEIR_E}\\
    I(t,\zeta_1,\zeta_2,\zeta_3) &= \sum^{\infty}_{i=0}I_i(t)\Phi(\zeta_1,\zeta_2,\zeta_3) = 
    \sum^{\infty}_{i=0}\sum^{\infty}_{j=0}\sum^{\infty}_{k=0}I_{ijk}(t)\Phi_i(\zeta_1)\Phi_j(\zeta_2)\Phi_k(\zeta_3)\label{eq:SEIR_I}\\
    R(t,\zeta_1,\zeta_2,\zeta_3) &= \sum^{\infty}_{i=0}R_i(t)\Phi(\zeta_1,\zeta_2,\zeta_3) =
    \sum^{\infty}_{i=0}\sum^{\infty}_{j=0}\sum^{\infty}_{k=0}R_{ijk}(t)\Phi_i(\zeta_1)\Phi_j(\zeta_2)\Phi_k(\zeta_3)\label{eq:SEIR_R}.
\end{align}
\end{subequations}
The polynomial $\Phi(\zeta_1,\zeta_2,\zeta_3)$ consists of all tensor product of Legendre polynomial of $\zeta_1$, $\zeta_2$ and $\zeta_3$. The ordinary differential equations can now be expressed as:
\begin{subequations}
\begin{align}
\sum^{\infty}_{i=0}\sum^{\infty}_{j=0}\sum^{\infty}_{k=0} \frac{dS_{ijk}(t)}{dt} \Phi_i(\zeta_1)\Phi_j(\zeta_2)\Phi_k(\zeta_3) = -\left(\beta_0 + \beta_1\zeta_1\right)...\nonumber\\
...\sum^{\infty}_{i=0}\sum^{\infty}_{j=0}\sum^{\infty}_{k=0}\sum^{\infty}_{m=0}\sum^{\infty}_{n=0}\sum^{\infty}_{q=0}S_{ijk}(t)\Phi_i(\zeta_1)\Phi_j(\zeta_2)\Phi_k(\zeta_3) \times I_{mnq}(t)\Phi_m(\zeta_1)\Phi_n(\zeta_2)\Phi_q(\zeta_3)\label{eq:SEIR_PCE_S}\\
\sum^{\infty}_{i=0}\sum^{\infty}_{j=0}\sum^{\infty}_{k=0} \frac{dE_{ijk}(t)}{dt} \Phi_i(\zeta_1)\Phi_j(\zeta_2)\Phi_k(\zeta_3) = \left(\beta_0 + \beta_1\zeta_1\right)...\nonumber\\
...\sum^{\infty}_{i=0}\sum^{\infty}_{j=0}\sum^{\infty}_{k=0}\sum^{\infty}_{m=0}\sum^{\infty}_{n=0}\sum^{\infty}_{q=0}S_{ijk}(t)\Phi_i(\zeta_1)\Phi_j(\zeta_2)\Phi_k(\zeta_3) \times I_{mnq}(t)\Phi_m(\zeta_1)\Phi_n(\zeta_2)\Phi_q(\zeta_3)...\nonumber\\
...- \left(\sigma_0 + \sigma_1\zeta_2\right) \sum^{\infty}_{i=0}\sum^{\infty}_{j=0}\sum^{\infty}_{k=0} E_{ijk}(t) \Phi_i(\zeta_1)\Phi_j(\zeta_2)\Phi_k(\zeta_3)\label{eq:SEIR_PCE_E}\\
\sum^{\infty}_{i=0}\sum^{\infty}_{j=0}\sum^{\infty}_{k=0} \frac{dI_{ijk}(t)}{dt} \Phi_i(\zeta_1)\Phi_j(\zeta_2)\Phi_k(\zeta_3) = \left(\sigma_0 + \sigma_1\zeta_2\right)...\nonumber\\
...\sum^{\infty}_{i=0}\sum^{\infty}_{j=0}\sum^{\infty}_{k=0}E_{ijk}(t)\Phi_i(\zeta_1)\Phi_j(\zeta_2)\Phi_k(\zeta_3) - \left(\gamma_0 + \gamma_1\zeta_3\right) \sum^{\infty}_{i=0}\sum^{\infty}_{j=0}\sum^{\infty}_{k=0} I_{ijk}(t) \Phi_i(\zeta_1)\Phi_j(\zeta_2)\Phi_k(\zeta_3)\label{eq:SEIR_PCE_I}\\
\sum^{\infty}_{i=0}\sum^{\infty}_{j=0}\sum^{\infty}_{k=0} \frac{dR_{ijk}(t)}{dt} \Phi_i(\zeta_1)\Phi_j(\zeta_2)\Phi_k(\zeta_3) = \left(\gamma_0 + \gamma_1\zeta_3\right) \sum^{\infty}_{i=0}\sum^{\infty}_{j=0}\sum^{\infty}_{k=0}I_{ijk}(t)\Phi_i(\zeta_1)\Phi_j(\zeta_2)\Phi_k(\zeta_3)\label{eq:SEIR_PCE_R}.
\end{align}
\end{subequations}
To approximate the original system shown in Eqns.~\eqref{eq:SEIR_original_S}-\eqref{eq:SEIR_original_R}, the 
Galerkin projection onto the subspace of the orthogonal basis functions formed by the tensor product of the Weiner-Askey basis functions in each random dimension leads to the equations:
%a polynomial of degree $P$ needs to be chosen. We perform Galerkin projections by multiplying with $\Phi_u(\zeta_1)$, $\Phi_v(\zeta_2)$ and $\Phi_w(\zeta_3)$ where \mbox{$u=0,\:1,\:2,\:...P$}, \mbox{$v=0,\:1,\:2,\:...(P-u)$} and \mbox{$w=0,\:1,\:2,\:...(P-u-v)$}, to keep even the mixed polynomial between $\zeta_1$, $\zeta_2$ and $\zeta_3$ at degree $P$. The closed form expression after a Galerkin projection can be written as follows:
\begin{subequations}
    \begin{align}
    \frac{dS_{uvw}}{dt} = -\frac{\beta_0}{\Psi_3}\sum^{P}_{i=0}\sum^{P-i}_{j=0}\sum^{P-i-j}_{k=0}\sum^{P}_{m=0}\sum^{P-m}_{n=0}\sum^{P-m-n}_{q=0}S_{ijk}(t)I_{mnq}(t)\Psi_2(1) ...\nonumber\\
    ...-\frac{\beta_1}{\Psi_3}\sum^{P}_{i=0}\sum^{P-i}_{j=0}\sum^{P-i-j}_{k=0}\sum^{P}_{m=0}\sum^{P-m}_{n=0}\sum^{P-m-n}_{q=0}S_{ijk}(t)I_{mnq}(t)\Psi_2(\zeta_1)\\
    \frac{dE_{uvw}}{dt} = \frac{\beta_0}{\Psi_3}\sum^{P}_{i=0}\sum^{P-i}_{j=0}\sum^{P-i-j}_{k=0}\sum^{P}_{m=0}\sum^{P-m}_{n=0}\sum^{P-m-n}_{q=0}S_{ijk}(t)I_{mnq}(t)\Psi_2(1) ...\nonumber\\
    ...+\frac{\beta_1}{\Psi_3}\sum^{P}_{i=0}\sum^{P-i}_{j=0}\sum^{P-i-j}_{k=0}\sum^{P}_{m=0}\sum^{P-m}_{n=0}\sum^{P-m-n}_{q=0}S_{ijk}(t)I_{mnq}(t)\Psi_2(\zeta_1)
    - \sigma_0E_{uvw}(t) - \frac{\sigma_1}{\Psi_3}\sum^{P}_{i=0}\sum^{P-i}_{j=0}\sum^{P-i-j}_{k=0}E_{ijk}(t)\Psi_1\left(\zeta_2\right)\\
    \frac{dI_{uvw}}{dt} = 
    \sigma_0E_{uvw}(t) + \frac{\sigma_1}{\Psi_3}\sum^{P}_{i=0}\sum^{P-i}_{j=0}\sum^{P-i-j}_{k=0}E_{ijk}(t)\Psi_1\left(\zeta_2\right) - \gamma_0I_{uvw}(t) - \frac{\gamma_1}{\Psi_3}\sum^{P}_{i=0}\sum^{P-i}_{j=0}\sum^{P-i-j}_{k=0}I_{ijk}(t)\Psi_1\left(\zeta_3\right)\\
    \frac{dR_{uvw}}{dt} = \gamma_0I_{uvw}(t) + \frac{\gamma_1}{\Psi_3}\sum^{P}_{i=0}\sum^{P-i}_{j=0}\sum^{P-i-j}_{k=0}I_{ijk}(t)\Psi_1\left(\zeta_3\right).
    \end{align}
\end{subequations}
$\Psi_1$, $\Psi_2$ and $\Psi_3$ are calculated as:
\begin{align}
\Psi_1\left(F\left(\zeta_1, \zeta_2, \zeta_3\right)\right) = \int_{\Omega_3}\int_{\Omega_2}\int_{\Omega_1} 
F\left(\zeta_1, \zeta_2, \zeta_3\right) \Phi_i(\zeta_1)\Phi_j(\zeta_2)\Phi_k(\zeta_3) \Phi_u(\zeta_1)\Phi_v(\zeta_2)  ...\nonumber\\
...\times \Phi_w(\zeta_3)
p_1\left(\zeta_1\right)
p_2\left(\zeta_2\right)
p_3\left(\zeta_3\right) d\zeta_1 d\zeta_2 d\zeta_3\\
\Psi_2\left(F\left(\zeta_1, \zeta_2, \zeta_3\right)\right) =  \int_{\Omega_3}\int_{\Omega_2}\int_{\Omega_1} 
F\left(\zeta_1, \zeta_2, \zeta_3\right) \Phi_i(\zeta_1)\Phi_j(\zeta_2)\Phi_k(\zeta_3)
\Phi_m(\zeta_1)\Phi_n(\zeta_2)\Phi_q(\zeta_3) ...\nonumber\\
...\times \Phi_u(\zeta_1)\Phi_v(\zeta_2)\Phi_w(\zeta_3) 
p_1\left(\zeta_1\right)
p_2\left(\zeta_2\right)
p_3\left(\zeta_3\right) d\zeta_1 d\zeta_2 d\zeta_3\\
\Psi_3 = \int_{\Omega_3}\int_{\Omega_2}\int_{\Omega_1} \left(\Phi_u\left(\zeta_1\right)\right)^2 \left(\Phi_v\left(\zeta_2\right)\right)^2
\left(\Phi_w\left(\zeta_3\right)\right)^2
p_1\left(\zeta_1\right)
p_2\left(\zeta_2\right)
p_3\left(\zeta_3\right) d\zeta_1 d\zeta_2 d\zeta_3 \label{eq:Psi_3}
\end{align}
and $p_1$, $p_2$ and $p_3$ are the pdfs of $\zeta_1$, $\zeta_2$ and $\zeta_3$ respectively. Note that $\Psi_3$ is simply a coefficients which is the result of the Galerkin projection of the left-hand side of Eqns.~\eqref{eq:SEIR_PCE_S}-\eqref{eq:SEIR_PCE_R} onto the orthogonal basis functions. 
%Hence, $\Psi_1$, $\Psi_2$ and $\Psi_3$ depend on which $u$, $v$ and $w$ are chosen.
%%%%%%%%%%%%%%%%%%%%%%%%%%%%%%%%%%%%%%%%%%%%%%%%%%%%%%%%
\subsubsection{Mean and variance}
For demonstrative purposes, consider a group of infected people. The mean (expected value) can be calculated as follows:
\begin{align}
\EX \left[I(t,\zeta_1,\zeta_2,\zeta_3)\right] 
&= \EX\left[\sum^{\infty}_{i=0}\sum^{\infty}_{j=0}\sum^{\infty}_{k=0}I_{ijk}(t)\Phi_1\left(\zeta_1\right)\left(\zeta_2\right)\left(\zeta_3\right)\right]\nonumber\\
&= \sum^{\infty}_{i=0}\sum^{\infty}_{j=0}\sum^{\infty}_{k=0}I_{ijk}(t)\int_{\Omega_3}\int_{\Omega_2}\int_{\Omega_1}\Phi_i\left(\zeta_1\right)\Phi_j\left(\zeta_2\right)\Phi_k\left(\zeta_3\right) p_1\left(\zeta_1\right)
p_2\left(\zeta_2\right)
p_3\left(\zeta_3\right) d\zeta_1 d\zeta_2 d\zeta_3 \nonumber\\
&= I_{000}(t)
\end{align}
and the variance as:
\begin{align}
\mathbb{V} \left[I(t,\zeta_1,\zeta_2,\zeta_3)\right] 
= \EX \left[I(t,\zeta_1,\zeta_2,\zeta_3)^2\right] - \left(\EX \left[I(t,\zeta_1,\zeta_2,\zeta_3)\right] \right)^2\nonumber\\
= \EX  \left[\sum^{\infty}_{i=0}\sum^{\infty}_{j=0}\sum^{\infty}_{k=0}\sum^{\infty}_{m=0}\sum^{\infty}_{n=0}\sum^{\infty}_{q=0}I_{ijk}(t)I_{mnq}(t)\Phi_i(\zeta_1)\Phi_j(\zeta_2)\Phi_k(\zeta_3)
\Phi_m(\zeta_1)\Phi_n(\zeta_2)\Phi_q(\zeta_3)\right] - I_{000}(t)^2\nonumber\\
= \sum^{\infty}_{i=0}\sum^{\infty}_{j=0}\sum^{\infty}_{k=0} \left(I_{ijk}\right)^2 \int_{\Omega_3}\int_{\Omega_2}\int_{\Omega_1}\left(\Phi_i\left(\zeta_1\right)\right)^2\left(\Phi_j\left(\zeta_2\right)\right)^2\left(\Phi_k\left(\zeta_3\right)\right)^2 ...\nonumber\\
....p_1\left(\zeta_1\right)
p_2\left(\zeta_2\right)
p_3\left(\zeta_3\right) d\zeta_1 d\zeta_2 d\zeta_3
- \left(I_{000}(t)\right)^2.
\end{align}
Furthermore we can simply calculate the conditional variance as shown in Eqn.~\eqref{eq:conditional_variance_1} and \eqref{eq:conditional_variance_2} which permit the calculation of the Sobol indices and Shapley effects.

\begin{figure}[H]
\centering
	\includegraphics[width=0.75\textwidth]{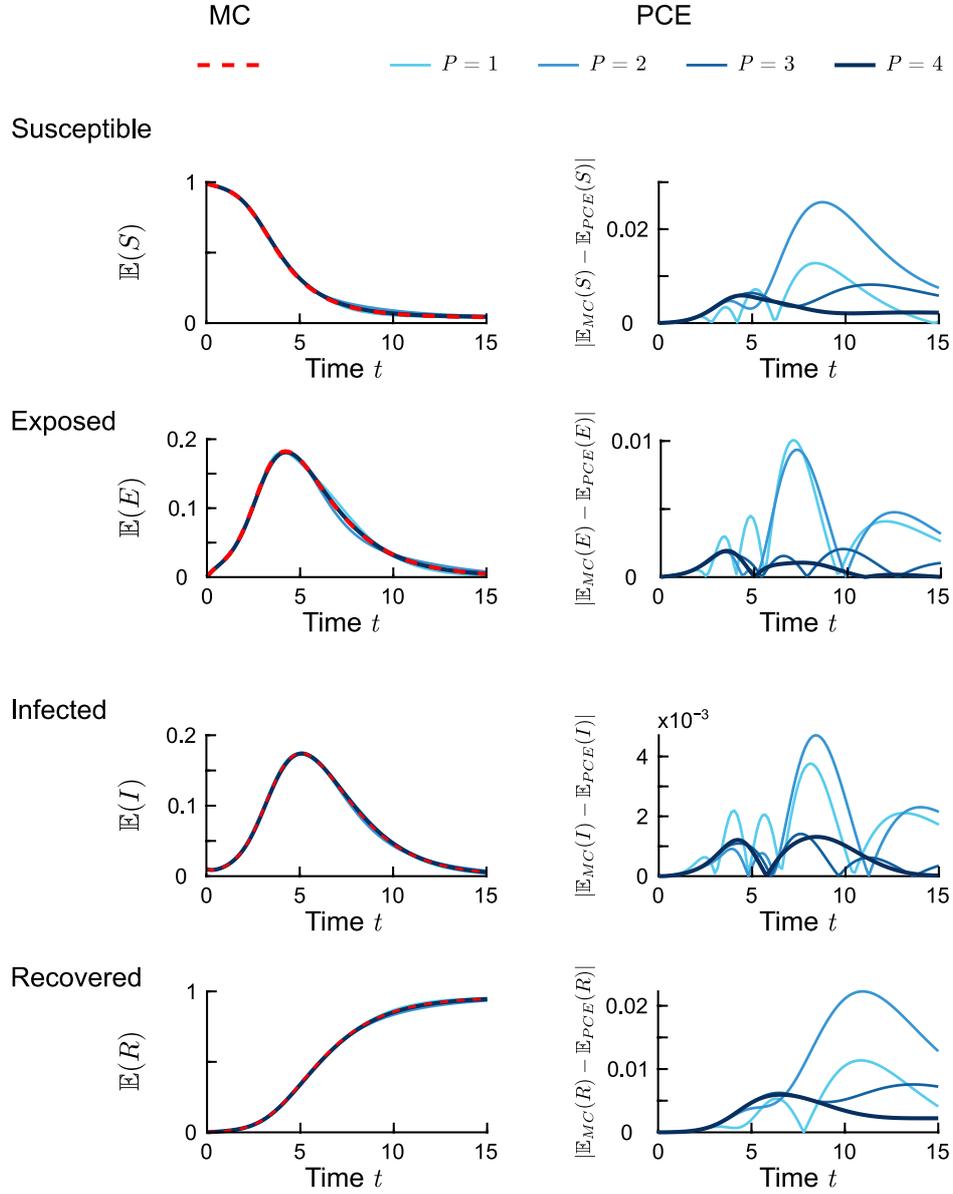}
	\caption{Expected value with Monte Carlo simulation ($5000$ samples) compared to the Polynomial Chaos Expansion for the susceptible, exposed, infected and recovered group (on the left). Absolute error of the expected value between Monte Carlo simulation ($5000$ samples) and the Polynomial Chaos Expansion (on the right).}
 \label{fig:SEIR_all_states_MC_vs_PCE_mean}
\end{figure}

\begin{figure}[H]
\centering
	\includegraphics[width=0.75\textwidth]{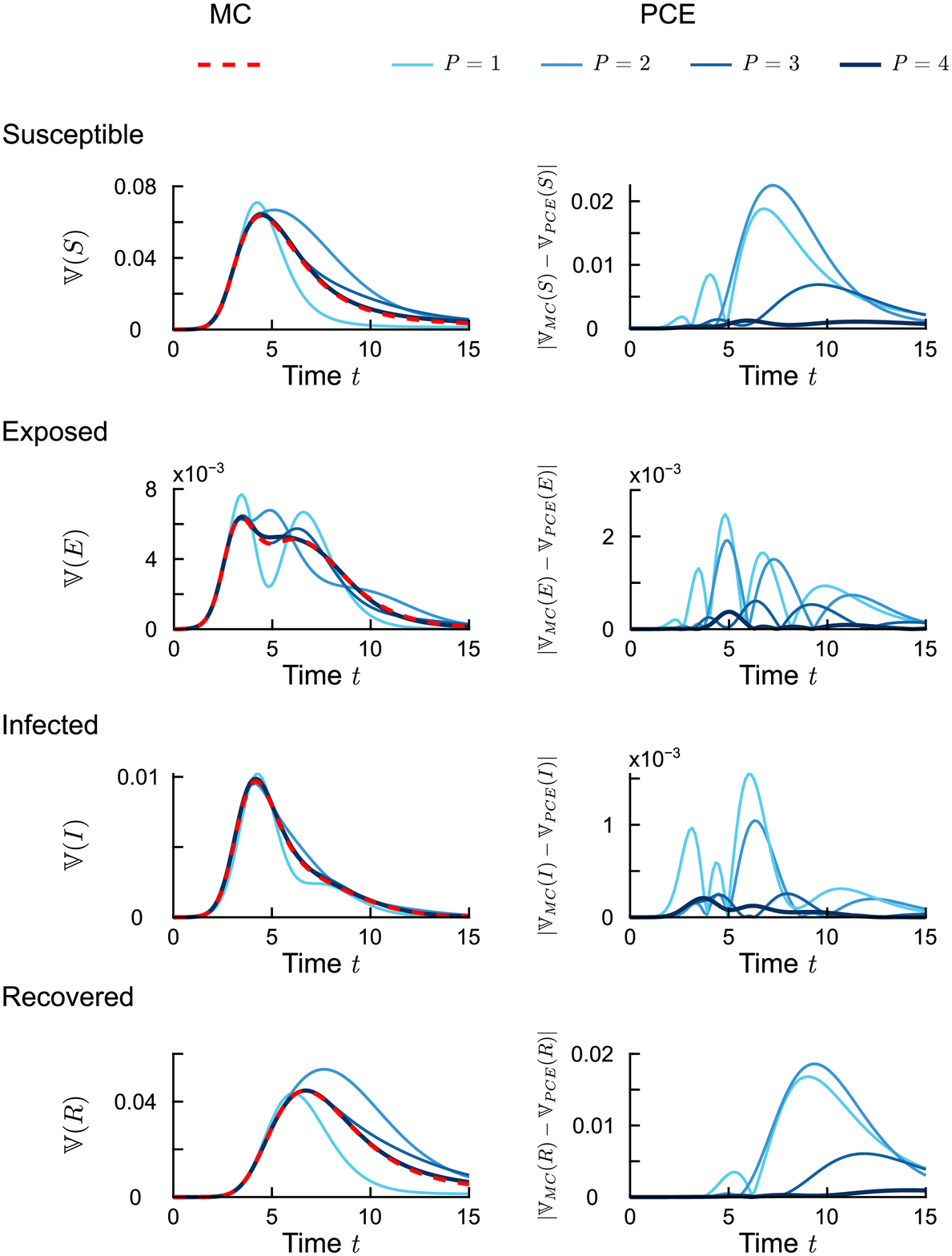}
	\caption{Variance with Monte Carlo simulation ($5000$ samples) compared to the Polynomial Chaos Expansion for the susceptible, exposed, infected and recovered group (on the left). Absolute error of the variance between Monte Carlo simulation ($5000$ samples) and the Polynomial Chaos Expansion (on the right).}
 \label{fig:SEIR_all_states_MC_vs_PCE_variance}
\end{figure}

\begin{figure}[H]
\centering
	\includegraphics[width=0.55\textwidth]{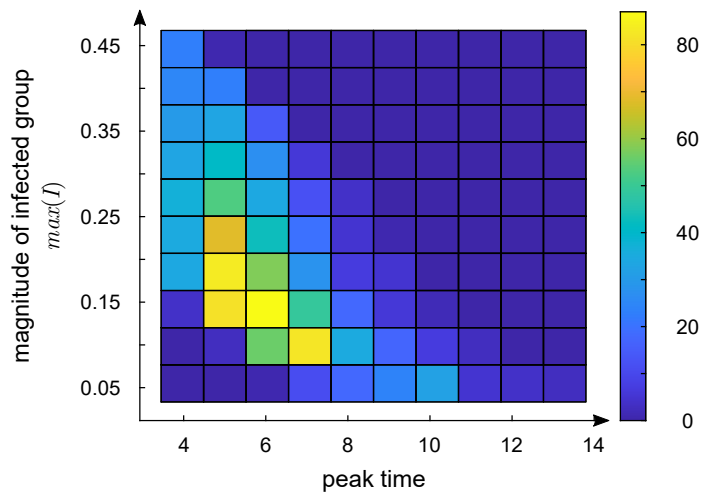}
	\caption{Heat map of the magnitude and peak time of the infected group.}
 \label{fig:Histogram_peak_time_and_magnitude}
\end{figure}

\begin{figure}[H]
\centering
	\includegraphics[width=\textwidth]{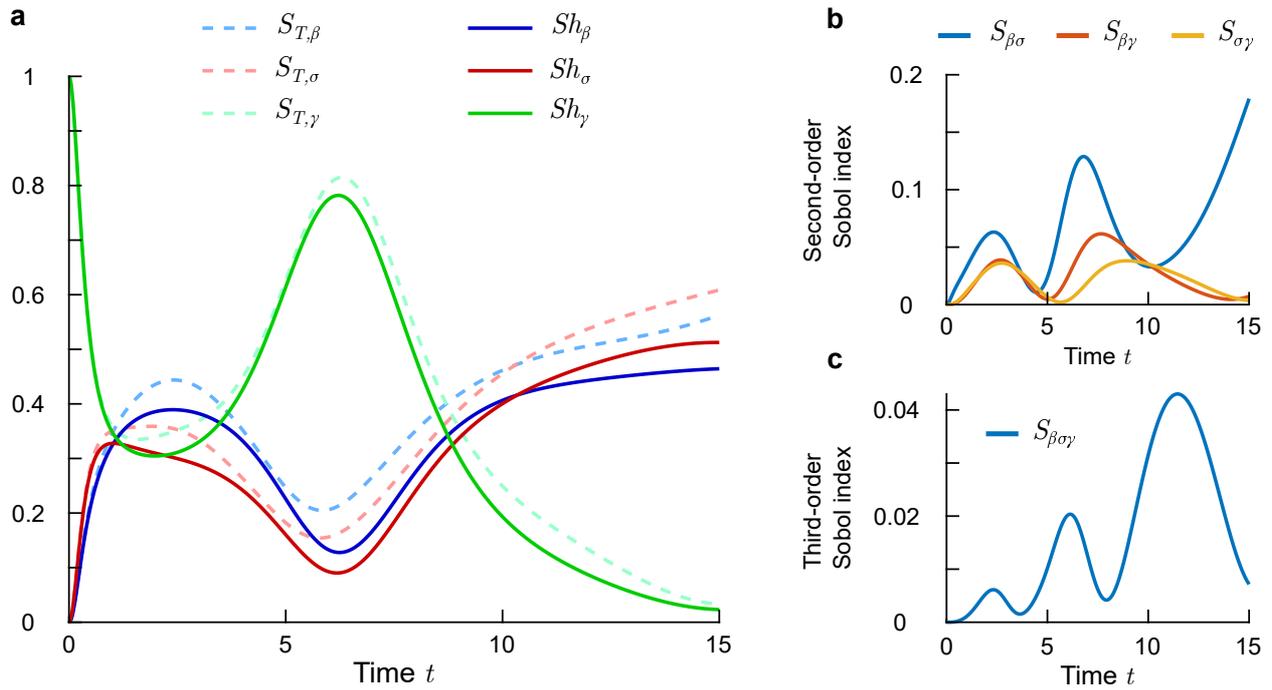}
	\caption{Importance of uncertain variables on the infected group derived from a Polynomial Chaos Expansion of degree $4$. a) Total Sobol indices (dashed lines) versus Shapley effects (solid lines). b) Second-order Sobol indices. c) Third-order Sobol indices.}
 \label{fig:SEIR_Infected_PCE_p4_Sobol_vs_Shapley}
\end{figure}

We start by gauging the approximation accuracy in evaluating the GSA metrics using PCE as compared to MC estimates. To compute the sensitivities with a variance-based approach, we only need the expected values and the variances, which is why we compare the MC to the PCE in terms of expected value and total variance. For the MC simulation, $5,000$ samples were chosen and considered sufficient to capture the uncertainties of the system and represent the states behavior. The PCE was computed until a degree of $P=4$. For the initial conditions it is assumed that $S_{000}=0.99$, $E_{000}=0$, $I_{000}=0.01$ and $R_{000}=0$. All other polynomial chaos states are $0$. 
Then we arbitrarily selected a simulation time of $15$ days. Figure~\ref{fig:SEIR_all_states_MC_vs_PCE_mean} illustrates the time variation of the expected value. The red dashed line corresponds to results computed with the MC method for the susceptible, exposed, infected and recovered group. The blue lines illustrate the results from the PCE where the increase in order of the PCE expansion corresponds to the increased depth of the blue color of the solid lines. From the left side of Figure~\ref{fig:SEIR_all_states_MC_vs_PCE_mean}, it is evident that the PCE approximations are converging to the results of the MC simulation. On the right side of Figure~\ref{fig:SEIR_all_states_MC_vs_PCE_mean} the absolute error over time between the MC and PCE is shown. The absolute error overall is fairly small; however, on average, a higher PCE degree yields a smaller error over time.
%%%%%%%%%%%%%%%%%%%%%%%%%%%%%%%
%%%%%%%%%%%%%%%%%%%%%%%%%%%%%%%
Figure~\ref{fig:SEIR_all_states_MC_vs_PCE_variance} illustrates via a red dashed line, the time-variation of the total variance generated by the MC simulations of the susceptible, exposed, infected and recovered group. The blue solid lines represent the simulation results of the PCE, with the higher PCE degrees being denoted by the increasingly darker blue colors. Compared to the expected value, the total variance shows a higher dependence on the degree of the PCE. The left side of Figure~\ref{fig:SEIR_all_states_MC_vs_PCE_variance} clearly illustrates that a low PC degree yields a poor approximation of the MC results; however, a degree of $P=4$ closely matches the MC results. The right side of Figure~\ref{fig:SEIR_all_states_MC_vs_PCE_variance} illustrates the absolute error between the MC and the PCE for all degrees and states. %Matching the observation of Figure~\ref{fig:SEIR_all_states_MC_vs_PCE_mean} the variance approximation error over time reduces with a higher PCE degree.
%%%%%%%%%%%%%%%%%%%%%%%%%%%%%%%
%%%%%%%%%%%%%%%%%%%%%%%%%%%%%%%
In disease forecasting, it is important to know when the infected group will peak, and at what magnitude. Given the PC-coefficients $I_{ijk}$ and basis functions $\Phi(\zeta_1,\zeta_2,\zeta_3)$, we can easily calculate the statistics of $I(t)$, by sampling the variables $\zeta_1$, $\zeta_2$ and $\zeta_3$ which are uniformly distributed over $[-1,1]$. The PC-coefficients must be calculated once and, by sampling the random variables, different trajectories of $I(t)$ over time can be synthesized. From these simulations, the largest magnitude of $I(t)$ and the peak time can be calculated and plotted in a 2D histogram.  Figure~\ref{fig:Histogram_peak_time_and_magnitude} illustrates a 2D heat-chart of the peak time and the magnitude of the infected group. It should be noted that the PCE is a lot faster in calculating such heat maps than a sample based approach such as classic MC methods. 
%The states of the system (Eqns.~\eqref{eq:SEIR_original_S}-\eqref{eq:SEIR_original_R}) don't have to be numerically integrated for all uncertain possibilities, because for the PCE approach the PC-coefficient get simply multiplied with the evaluated basis functions. 
In Figure~\ref{fig:Histogram_peak_time_and_magnitude} the peak time is most likely around time unit $5$, with a magnitude of approximately $0.18$, which matches the observation from the expected value in Figure~\ref{fig:SEIR_all_states_MC_vs_PCE_mean}.
%%%%%%%%%%%%%%%%%%%%%%%%%%%%%%%
%%%%%%%%%%%%%%%%%%%%%%%%%%%%%%%
The remaining question is: Do Shapley effects differ from Sobol indices for dynamical system as they did for the static equations which are presented in subsection \ref{subsec:Ishigami} and \ref{subsec:user_defined}? Table~\ref{table:Ishigami_Sobol_vs_Shapley_analytical_weights} indicates that Shapley is assigning less weight on higher order Sobol indices than when calculating the total Sobol indices. Figure~\ref{fig:SEIR_Infected_PCE_p4_Sobol_vs_Shapley} illustrates the importance via Sobol and Shapley of the three uncertain variables $\beta$, $\sigma$ and $\gamma$ over time on the infected group. The solid lines represent the Shapley effects while the dashed lines are the total Sobol indices. In subfigure a) initially the variable $\gamma$ has a high importance for the infected group. With the given initial conditions and looking at the Eqns.~\eqref{eq:SEIR_original_S}-\eqref{eq:SEIR_original_R}, it can be noted the initial high importance of $\gamma$ is justified because the infected group reduces its value by $\gamma$ to the recovered group. At the initial time the variables $\beta$ and $\sigma$ have no impact on the infected group and are therefore ranked very low because the exposed group is initialized with $0$. As time goes on the importance of $\gamma$ drops and the importance of $\beta$ and $\sigma$ rises, while $\beta$ is higher valued than $\sigma$. This occurs because $\beta$ provides the growth of the exposed group which then goes over to the infected group. Looking back at Figure~\ref{fig:SEIR_all_states_MC_vs_PCE_mean}, the expected value peaks at time instant $6$, and from Eqn.~\eqref{eq:SEIR_original_I}, it is obvious that if $I$ becomes large, the variable $\gamma$ has a large impact. This is captured in Figure~\ref{fig:SEIR_Infected_PCE_p4_Sobol_vs_Shapley} a) as well. When the infected group is reduced, the importance of $\gamma$ shrinks and $\beta$ and $\sigma$ become again more important because they provide the growth of $I$.

\begin{figure}[H]
\centering
	\includegraphics[width=\textwidth]{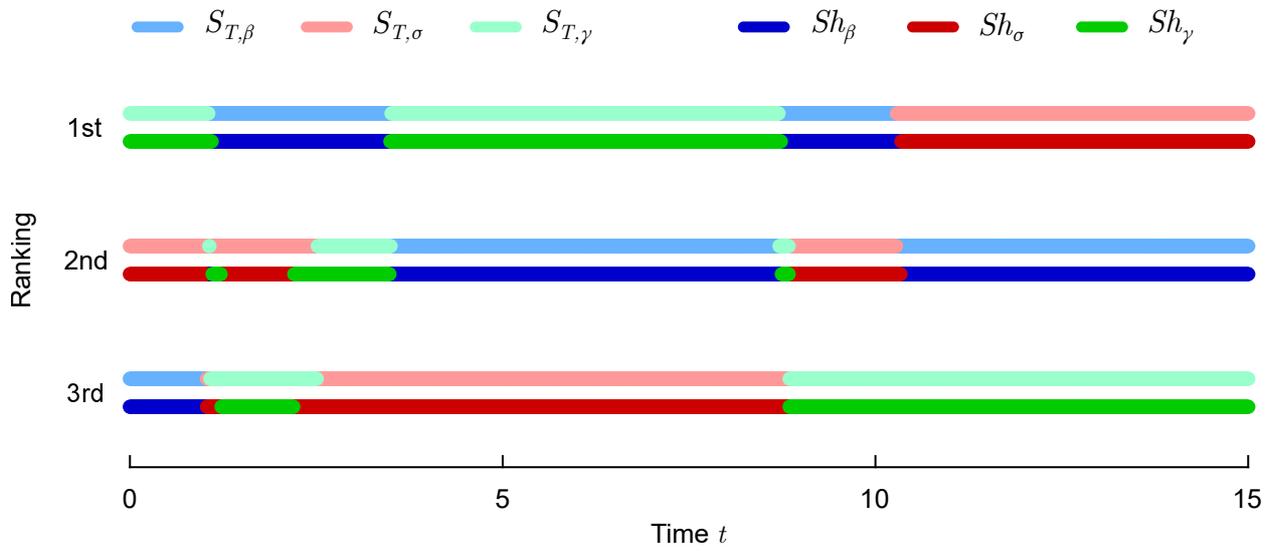}
	\caption{Ranking of the uncertain parameters in the SEIR model according to the total Sobol indices and Shapley effects.}
 \label{fig:SEIR_Infected_PCE_p4_Sobol_vs_Shapley_ranking}
\end{figure}

\begin{figure}[H]
\centering
	\includegraphics[width=\textwidth]{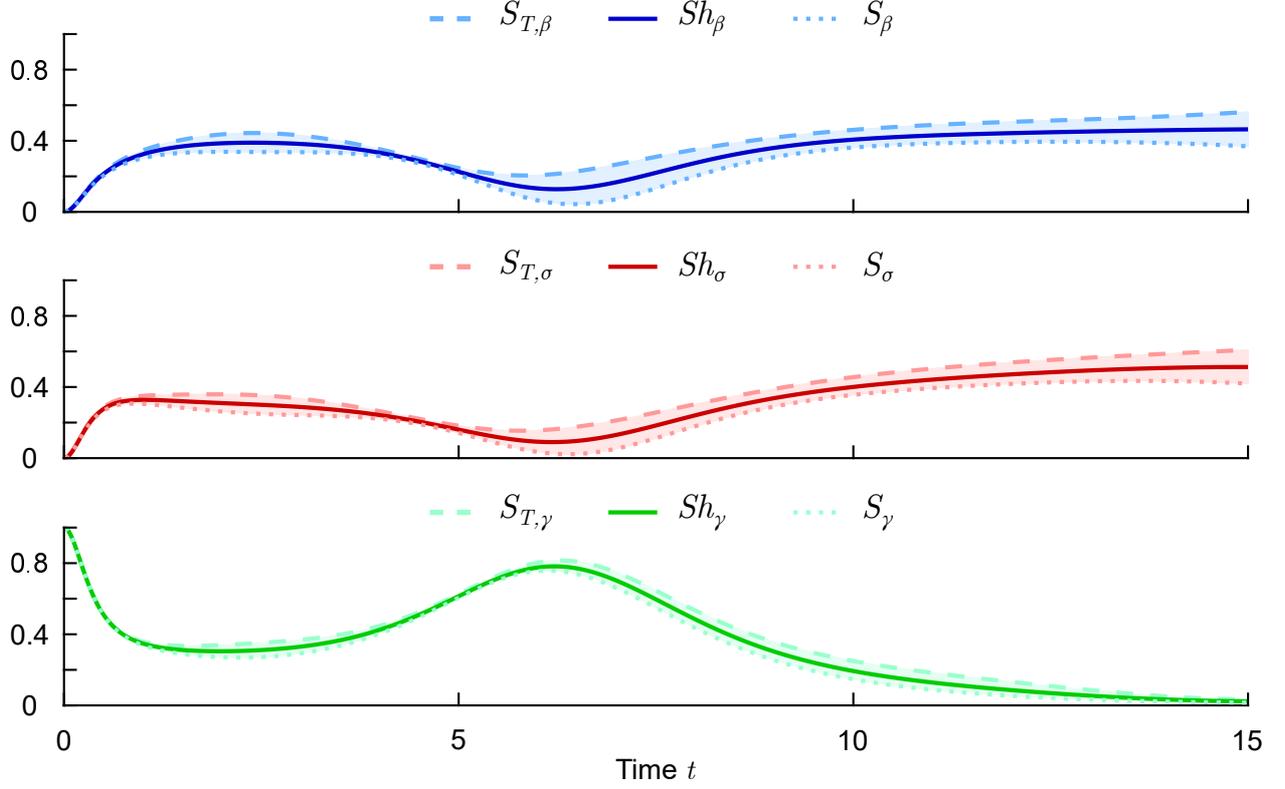}
	\caption{Sandwiching of Shapley effects between Total Sobol indices (upper bound) and first-order Sobol indices (lower bound) for the Susceptible-Exposed-Infected-Recovered disease model.}
 \label{fig:SEIR_Infected_PCE_p4_Sandwiching}
\end{figure}

Figure~\ref{fig:SEIR_Infected_PCE_p4_Sobol_vs_Shapley} illustrates the time-variation of a) the total Sobol indices and Shapley effects are plotted on the same scale, where we note that Shapley effects sum up to $1$ for all times while this is not true for the total Sobol indices. However, Figure~\ref{fig:SEIR_Infected_PCE_p4_Sobol_vs_Shapley} a) also illustrates the relative importance of all the uncertain variables for both approaches. To answer the question of why the total Sobol indices and Shapley effects differ, we need to look at the higher order Sobol indices, which are shown in Figure~\ref{fig:SEIR_Infected_PCE_p4_Sobol_vs_Shapley} b) and c). Their different weighting results in divergence of the Shapley effects from the total Sobol indices. It can be seen that especially when these higher Sobol indices peak that the total Sobol indices don't coincide with the Shapley effects (for instance around time $6$) and when they are nearly $0$ (around time $5$) that the Shapley effects and total Sobol indices are almost the same. However, it is rather difficult to conclude a ``ranking'' order from Figure~\ref{fig:SEIR_Infected_PCE_p4_Sobol_vs_Shapley} a). 
%%%%%%%%%%%%%%%%%%%%%%%%%%%%%%%
%%%%%%%%%%%%%%%%%%%%%%%%%%%%%%%
Therefore, we introduce Figure~\ref{fig:SEIR_Infected_PCE_p4_Sobol_vs_Shapley_ranking} where the colors blue, red, and green represent the uncertain variables $\beta$, $\sigma$ and $\gamma$ respectively. The ranking of each uncertain variables are provided between the initial and final time in an intuitive ranking scheme of $1$st, $2$nd and $3$rd. A significant ranking difference of the uncertain variables can be seen between the times $1$ to $4$, which is just before the expected value of the infected group peaks. This information might be very helpful for entities that are trying to take certain measures to slow down the growth of infections in a pandemic. The total Sobol indices rank variable $\sigma$ at $2$nd place until almost time instant $3$ while Shapley suggests that the rank switch between $\sigma$ and $\gamma$ is happening earlier (around time $2.5$). After the expected value of the infected group peaks, the total Sobol indices and the Shapley effects are similar.
%%%%%%%%%%%%%%%%%%%%%%%%%%%%%%%
%%%%%%%%%%%%%%%%%%%%%%%%%%%%%%%
Owen~\cite{Owen.2014} pointed out that the Shapley effect is ``sandwiched" in between the first-order Sobol index and the total order Sobol index. This can be seen in Figure~\ref{fig:SEIR_Infected_PCE_p4_Sandwiching} for the uncertain variables $\beta$, $\sigma$ and $\gamma$, where the dashed, solid and dotted lines represent the total Sobol indices, Shapley effects, and the first-order Sobol indices respectively. The sandwiching size is particularly large for $\beta$ and $\sigma$ around time instant $6$, when the higher order Sobol indices peak as indicated in Figure~\ref{fig:SEIR_Infected_PCE_p4_Sobol_vs_Shapley} b) and c). In contrast, the width of the sandwich is negligible around time instant $5$, which is when the higher order Sobol indices are nearly $0$.
%%%%%%%%%%%%%%%%%%%%%%%%%%%%%%%
%%%%%%%%%%%%%%%%%%%%%%%%%%%%%%%
%%%%%%%%%%%%%%%%%%%%%%%%%%%%%%%%%%%%%%%%%%%%%%%%%%%%%%%%
\subsection{Type 1 diabetes Bergman model}
\begin{figure}[H]
\centering
	\includegraphics[width=0.75\textwidth]{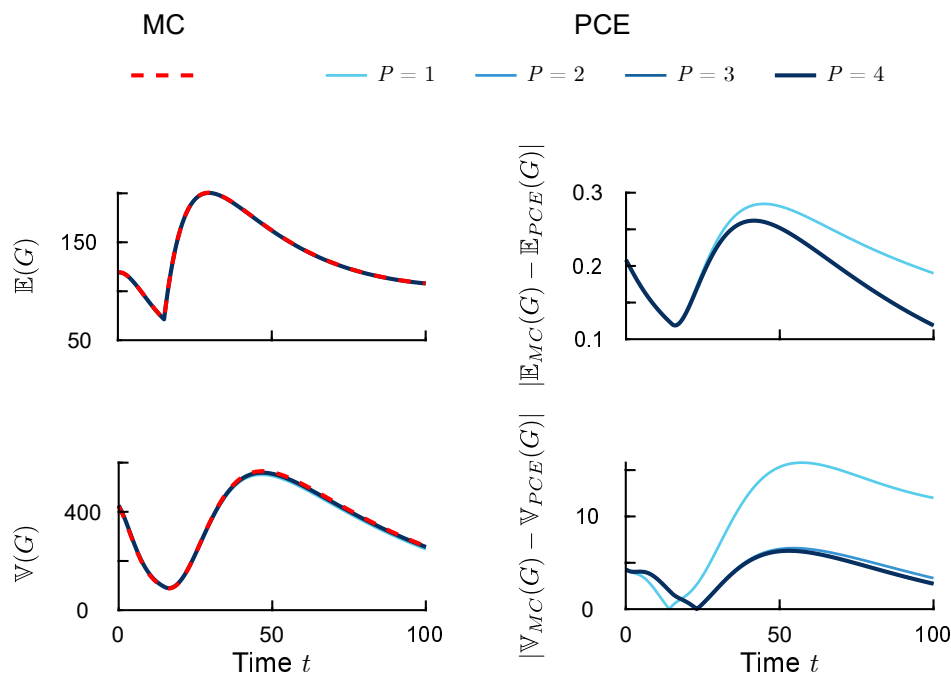}
	\caption{Expected value and variance with Monte Carlo simulation (10000 samples) compared to the Polynomial Chaos Expansion for blood glucose level (on the left). Absolute error of the expected value and variance between Monte Carlo simulation (10000 samples) and the Polynomial Chaos Expansion (on the right).}
 \label{fig:Bergman_Glucose_MC_vs_PCE_mean_and_var}
\end{figure}

\begin{figure}[H]
\centering
	\includegraphics[width=\textwidth]{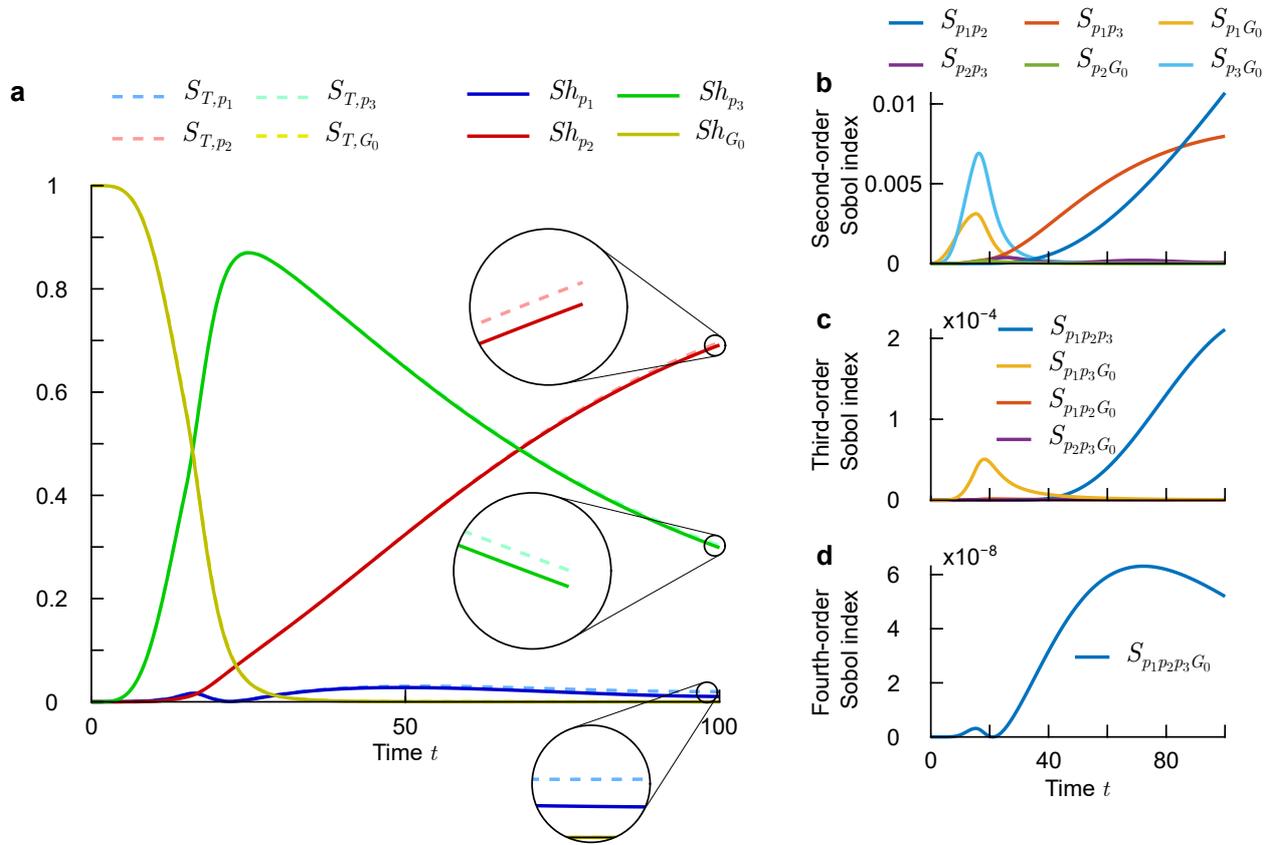}
	\caption{Importance of uncertain variables on the blood glucose level derived from a Polynomial Chaos Expansion of degree $4$. a) Total Sobol indices (dashed lines) versus Shapley effects (solid lines). b) Second-order Sobol indices. c) Third-order Sobol indices. d) Fourth-order Sobol indices.}
 \label{fig:Bergman_Glucose_PCE_p4_Sobol_vs_Shapley}
\end{figure}

\begin{figure}[H]
\centering
	\includegraphics[width=\textwidth]{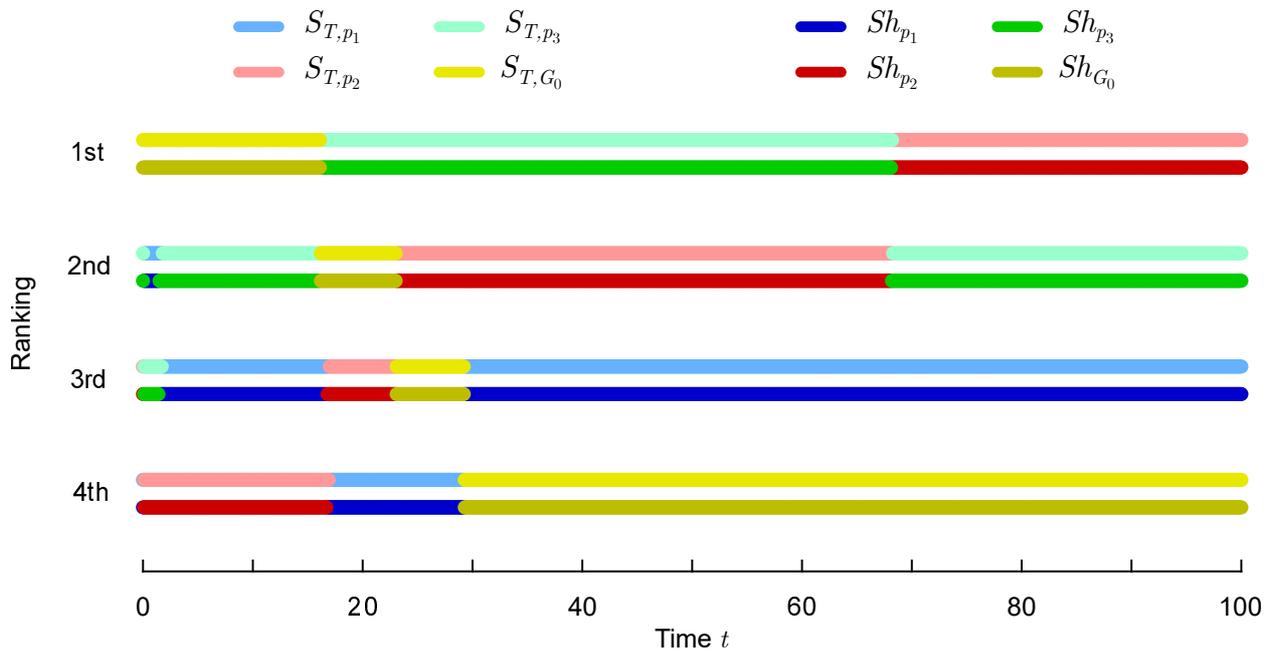}
	\caption{Ranking of the uncertain parameters in the Bergman model according to the total Sobol indices and Shapley effects.}
 \label{fig:Bergman_Glucose_PCE_p4_Sobol_vs_Shapley_ranking}
\end{figure}

\begin{figure}[H]
\centering
	\includegraphics[width=\textwidth]{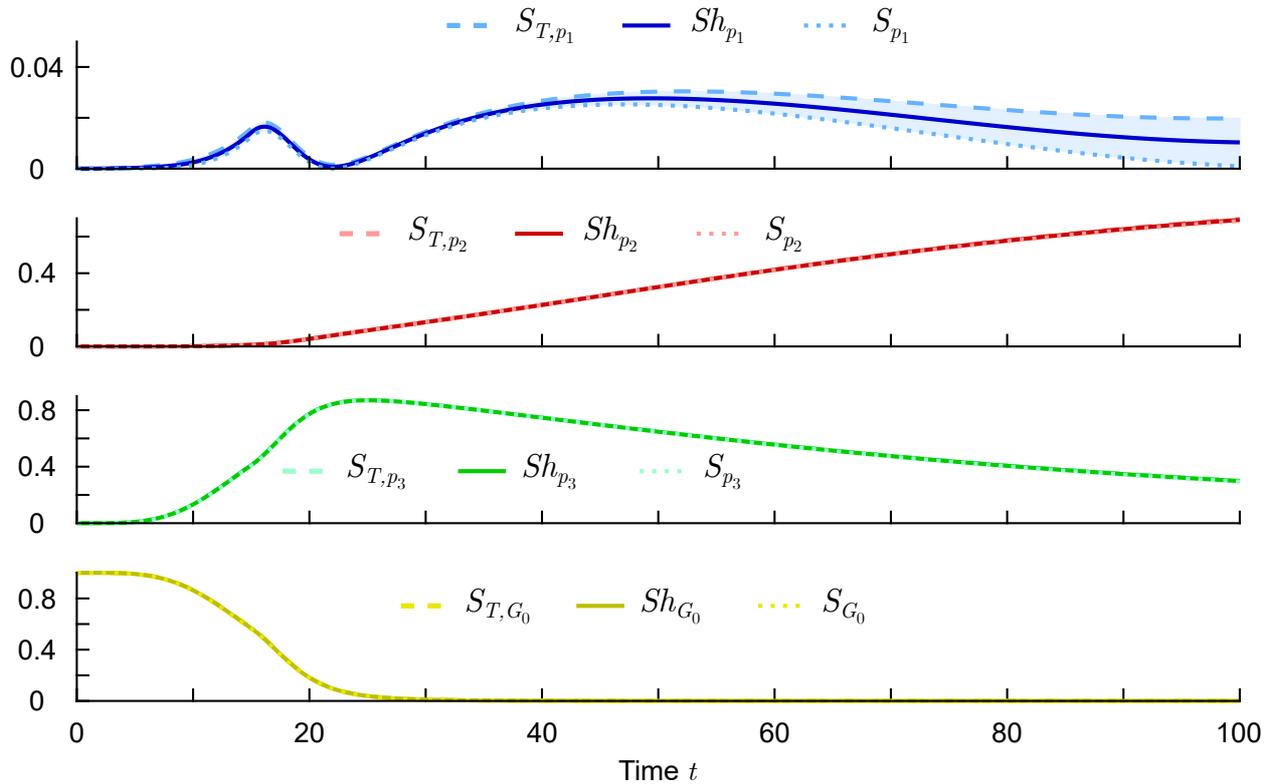}
	\caption{Sandwiching of Shapley effects between Total Sobol indices (upper bound) and first-order Sobol indices (lower bound) for the  Bergman model.}
 \label{fig:Bergman_Glucose_PCE_p4_Sobol_vs_Shapley_Sandwich}
\end{figure}

A popular model to study type 1 Diabetes (T1D) is the minimal Bergman model~\cite{Nandi.2020}:
\begin{subequations}
    \begin{align}
        \dot{G}(t) &= -\left(X(t)+p_1\right)G(t) + p_1G_b + D(t)\label{eq:Bergman_G}\\
        \dot{X}(t) &= -p_2X(t) + p_3\left(I(t)-I_b\right)\label{eq:Bergman_X}\\
        \dot{I}(t) &= -p_4\left(I(t)-I_b\right) + U(t) \label{eq:Bergman_I}
    \end{align}
\end{subequations}
where $G$ is the blood glucose level, $X$ is the intermediate state and $I$ is the insulin concentration. The variable $D(t)$ is the meal disturbance term and is assumed to be the Fisher model~\cite{Fisher.1991}:
\begin{align}
    D(t) = 
    \begin{cases}
        0 & \:\:\: t< t_m\\
        Be^{-d\left(t-t_m\right)} & \:\:\: t\geq t_m
    \end{cases}
\end{align}
where $B$ stands for the meal quantity and is assumed to be $B=28.98$, calculated from a meal of $45$ gm Carbohydrates (CHO) and the meal time is given as $t_m=15$ min \cite{Nandi.2020}. This simulation considers an insulin bolus at $t=0$. For the simulation the insulin bolus is represented by the time-varying term $U(t)$ in Eqn.~\eqref{eq:Bergman_I}. The input $U(t)$ is given by:
\begin{align}
    U(t) = 
    \begin{cases}
        \frac{1000 \cdot CHO (\text{in g})}{CR \cdot V_i} & \:\:\: 0 < t < 1\\
        0 &\:\:\: 1 < t
    \end{cases}
\end{align}
where $CR$ is the insulin-to-carb ratio and $V_i$ is the distribution volume-to-insulin. For a $45$ gm meal, $CR=18.477$ and $V_i=12$, and therefore $U(0<t<1)=202.96$ mU/L. The uncertainties lie in the parameters $p_1$, $p_2$, $p_3$ and the initial glucose level $G_0$. We assume that they are uniformly distributed as follows \cite{Nandi.2020}:
\begin{subequations}
    \begin{align}
    p_1 & \sim U(0.0201, 0.0373)\\
    p_2 & \sim U(0.0198, 0.0368)\\
    p_3 & \sim U(3.525e^{-5}, 6.545e^{-5})\\
    G_0 & \sim U(83.43, 154.93).
    \end{align}
\end{subequations}
When applying the PCE it should be noted that initial blood glucose is calculated by the Galerkin projections. Note that the initial blood glucose levels are divided by the coefficient which results from the Galerkin projection on the state equations. For the MC to PCE comparison, $10,000$ samples for the MC are chosen. The system is further initialized with $X_{0000}=0$ and $I_{0000}=15.3872$, while the rest of the expanded states of $X$ and $I$ are initialized to $0$. Figure~\ref{fig:Bergman_Glucose_MC_vs_PCE_mean_and_var} illustrates the expected value and total variance of the blood glucose level. A higher PCE degree leads to a smaller error between the MC and PCE approach. On the right side of Figure~\ref{fig:Bergman_Glucose_MC_vs_PCE_mean_and_var}, it appears that even with a higher degree of the PCE, a small error remains. 
%It might be because $10000$ samples for the MC are not completely representing the system's uncertain behavior.
%%%%%%%%%%%%%%%%%%%%%%%%%%%%%%%
%%%%%%%%%%%%%%%%%%%%%%%%%%%%%%%
Figure~\ref{fig:Bergman_Glucose_PCE_p4_Sobol_vs_Shapley} a) shows the total Sobol indices and Shapley effects over time of $p_1$, $p_2$, $p_3$, and $G_0$ with respect to the blood glucose level. As expected, the initial blood glucose level is the most important, meanwhile, as time progresses, the importance of $p_3$ peaks at approximately $25$ minutes, and then falls continuously. This is due to the insulin input in Eqn.~\eqref{eq:Bergman_I} which makes the insulin level rise, and therefore the uncertain parameters $p_3$ in Eqn.~\eqref{eq:Bergman_X} becomes more important. It can be observed that the importance of variable $p_2$ increases over time while the importance of variable $p_1$ is nominally the least important factor over all time. However, Figure~\ref{fig:Bergman_Glucose_PCE_p4_Sobol_vs_Shapley} a) also shows that, in the end, the Shapley effects and total Sobol indices diverge due to the higher Sobol indices as shown in Figure~\ref{fig:Bergman_Glucose_PCE_p4_Sobol_vs_Shapley} b), c), and d) which are relatively small except towards the end of the simulation time. The insets in Figure~\ref{fig:Bergman_Glucose_PCE_p4_Sobol_vs_Shapley} a) illustrate the divergence; however, this does not change the rank order of the system.
%%%%%%%%%%%%%%%%%%%%%%%%%%%%%%%
%%%%%%%%%%%%%%%%%%%%%%%%%%%%%%%
Figure~\ref{fig:Bergman_Glucose_PCE_p4_Sobol_vs_Shapley_ranking} illustrates the time-varying rank order of the uncertain variables with respect to the blood glucose level. The Shapley effects provides exactly the same answer as the total Sobol indices. This is due to the small values of the higher order Sobol indices which means they have a small impact on the calculation of the total Sobol indices and Shapley effects. Hence, both methods provide similar rankings. If we compare the higher order Sobol indices from the SEIR model as shown in Figure~\ref{fig:SEIR_Infected_PCE_p4_Sobol_vs_Shapley} with ones in Figure~\ref{fig:Bergman_Glucose_PCE_p4_Sobol_vs_Shapley} we can see that the magnitude of the second-order Sobol indices in the SEIR model are one order higher than in the Bergman Type 1 Diabetes model. The third-order Sobol indices are even two orders higher.
%%%%%%%%%%%%%%%%%%%%%%%%%%%%%%%
%%%%%%%%%%%%%%%%%%%%%%%%%%%%%%%
Figure~\ref{fig:Bergman_Glucose_PCE_p4_Sobol_vs_Shapley_Sandwich} presents the sandwiching effect of the First and Total Sobol indices on the uncertain variables $p_1$, $p_2$, $p_3$ and $G_0$. As shown in Figure~\ref{fig:Bergman_Glucose_PCE_p4_Sobol_vs_Shapley}, the total Sobol indices do not differ too much from the Shapley effects except for $p_1$ during the end of the simulation time. Therefore, a wide sandwiching effect can be mostly observed over the latter part of the simulation for $p_1$.
%%%%%%%%%%%%%%%%%%%%%%%%%%%%%%%
%%%%%%%%%%%%%%%%%%%%%%%%%%%%%%%
%%%%%%%%%%%%%%%%%%%%%%%%%%%%%%%%%%%%%%%%%%%%%%%%%%%%%%%%
%%%%%%%%%%%%%%%%%%%%%%%%%%%%%%%%%%%%%%%%%%%%%%%%%%%%%%%%
\section{Conclusion}
\label{sec:conclusion}
This paper proposes to use the Shapley value concept as a variance based global sensitivity metric and compares it to the well established variance based Sobol indices. The Shapley effects are evaluated for the Ishigami benchmark problem in addition to a polynomial model. It is noteworthy that the Shapley effects provide a ranking which is in conflict with the Total Sobol indices for the Ishigami function - a fact which has been illustrated using the Borgonovo delta function and other statistical distances. To address the issue of the computational burden in determining the Shapley effects, a polynomial chaos expansion has been proposed to be used to efficiently evaluate the total and conditional variances which are required to determine both the Sobol indices and Shapley effects. In effect, once a polynomial chaos model has been determined, the Sobol indices and the Shapley effects can be algebraically evaluated. The polynomial chaos approach has been used to illustrate the determination of Shapley effects of uncertain variables for an SEIR model and the Bergman Type 1 diabetes model.

The comparison between the total Sobol indices, Shapley effects and the Borgonovo metric illustrated the potential of ranking uncertain variables using Shapley effects. Apart from diabetes research, we aim to work on controller design techniques where the Shapley value concept could be used to desensitize controllers to uncertainties. In doing so, a mapping from polynomial chaos coefficients to Shapley effects will enable a low computational cost approach for robust controller design.
%%%%%%%%%%%%%%%%%%%%%%%%%%%%%%%%%%%%%%%%%%%%%%%%%%%%%%%%
%%%%%%%%%%%%%%%%%%%%%%%%%%%%%%%%%%%%%%%%%%%%%%%%%%%%%%%%
\section*{Acknowledgment}
The authors acknowledge the support of this work by the US National Science Foundation through CMMI Award number 2021710. 
\vspace{-0.1in}
\bibliographystyle{elsarticle-num}
\bibliography{ref} 
\noindent\rule[0.5ex]{\linewidth}{1pt}

\newpage
\onecolumn
\renewcommand{\theequation}{A-\arabic{equation}}     
% redefine the command that creates the equation no.    
\setcounter{equation}{0}

\end{document}